\documentclass{aa}
\usepackage{longtable}
\usepackage[varg]{txfonts}
\usepackage{epsfig}
\usepackage{natbib}
\bibpunct{(}{)}{;}{a}{}{,} 

\def\0{\hspace*{0.5em}}
\begin{document}


\title{Discovery of the magnetic field in the pulsating B star $\beta$ Cephei\thanks{Based on observations obtained using the MuSiCoS spectropolarimeter at
the Observatoire du Pic du Midi and
by the International Ultraviolet Explorer, which were collected at NASA Goddard Space Flight Center
and Villafranca Satellite Tracking Station of the European Space Agency.}}

\author{H.F. Henrichs\inst{1}
\and J.A. de Jong\inst{1,8}
    \and E. Verdugo\inst{2}
    \and R.S. Schnerr\inst{1}
    \and C. Neiner\inst{3}
    \and J.-F. Donati\inst{4}
    \and C. Catala\inst{5}
    \and S.L.S. Shorlin\inst{6}\and G.A. Wade\inst{7}
    \and P.M. Veen\inst{1}\and J.S. Nichols\inst{9}
    \and E.M.F. Damen\inst{1}
    \and A. Talavera\inst{10}
    \and G.M. Hill\inst{11}
    \and L. Kaper\inst{1}
    \and A.M. Tijani\inst{1}
    \and V.C. Geers\inst{1}
    \and K. Wiersema\inst{1}
    \and B. Plaggenborg\inst{1}
    \and K.L.J. Rygl\inst{1}
}

\institute{
Astronomical Institute "Anton Pannekoek", University of Amsterdam,
Science Park 904, 1098 XH Amsterdam, Netherlands
\and European Space Agency (ESAC), PO Box 78, 28691 Villanueva de la Ca\~{n}ada (Madrid), Spain
\and LESIA, Observatoire de Paris, CNRS
UMR 8109, UPMC, Universit\'e Paris Diderot; 5 place Jules Janssen, 92190
Meudon, France
\and Observatoire Midi-Pyr\'en\'ees, 14 Avenue Edouard Belin, 31400 Toulouse, France
\and LESIA, Observatoire de Paris, CNRS, UPMC, Universit\'{e} Paris Diderot, Place Jules Janssen, 92190 Meudon, France
\and Physics and Astronomy Department, The University of Western Ontario,
  London, Ontario, Canada N6A 3K7
\and Department of Physics, Royal Military College of Canada, P.O. Box 17000, Kingston, Ontario, Canada K7K 7B4
\and Max Planck Institute for Extraterrestrial Physics, Giessenbachstra\ss e, 85748 Garching bei M\"{u}nchen, Germany
\and Harvard-Smithsonian Center for Astrophysics, 60 Garden Str., Cambridge, MA
 02138, U.S.A.
\and European Space Astronomy Center, Villanueva de la Ca\~{n}ada, E-28691, Madrid, Spain
\and W.~M. Keck Observatory, 65-1120 Mamalahoa Highway, Kamuela, HI 96743, U.S.A.
}

\offprints{H.F. Henrichs, \email{h.f.henrichs@uva.nl}}

\date{Received date 27/03/2013 / Accepted date 22/04/2013}

\abstract
{Although the star itself is not helium enriched, the periodicity and the variability in the
UV wind lines of the pulsating B1~IV star $\beta$~Cephei are similar to
what is observed in magnetic helium-peculiar B stars, suggesting that $\beta$~Cep is magnetic.}
{We searched for a magnetic field using high-resolution spectropolarimetry. From UV spectroscopy, we analysed the wind variability and investigated the correlation with the magnetic data.} 
{Using 130 time-resolved circular polarisation spectra that were
obtained with the MuSiCoS \'{e}chelle spectropolarimeter at the 2m Telescope
Bernard Lyot from 1998, when $\beta$ Cep was discovered to be magnetic, until 2005, we applied the least-squares deconvolution method on the Stokes $V$ spectra and derived the longitudinal component of the integrated magnetic field over the visible hemisphere of the star. We performed a period analysis on the magnetic data and on equivalent-width measurements of UV wind lines obtained over 17 years.  We also analysed the short- and long-term radial velocity variations, which are due to the pulsations and the 90-year binary motion, respectively.}
{$\beta$ Cep hosts a sinusoidally varying magnetic field with an amplitude 97 $\pm$ 4 G and an average value $-6$ $\pm$ 3 G.
From the UV wind line variability, we derive a period of 12.00075(11) days,
which is the rotation period of the star, and is compatible with the observed magnetic modulation. Phases of maximum and minimum field match those of maximum emission in the UV wind lines, strongly supporting an oblique
magnetic-rotator model. We discuss the magnetic behaviour as a function of pulsation behaviour and UV line variability.}
{This paper presents the analysis of the first confirmed detection of a dipolar magnetic
field in an upper main-sequence pulsating star. Maximum wind absorption originates in the magnetic equatorial plane. Maximum emission occurs when the magnetic north pole points to the Earth.
Radial velocities agree with the $\sim$90-year orbit around its Be-star binary companion.}

\maketitle

\titlerunning{The magnetic field of $\beta$~Cephei}

\authorrunning{H.~F. Henrichs, et al.}

\section{Introduction}

The star \object{$\beta$~Cephei} (\object{HR 8238}, \object{HD 205021}, \object{HIP 106032}, \object{WDS 21287+7034}) was classified as
spectral type B1~III by \cite{lesh:1968}, but earlier references suggest that it may be type B2~III
or B1~IV. \cite{morel:2006} assigned it a revised spectral type of B1~Vevar.
The Be status of this star was dismissed by \cite{schnerr:2006e}, who unambiguously demonstrated by spectroastrometry that the intermittent H$\alpha$ emission often encountered in spectra of $\beta$ Cep
originates in the companion star (with spectral type B6--8) that is 3.4-magnitude fainter and mostly only speckle-resolved.
This companion was discovered with the 200-inch Hale telescope by \cite{gezari:1972}, and its
orbit with an approximate period of
90 years was subsequently determined by \cite{pigulski:1992}.  More properties of this companion have been studied by \cite{wheelwright:2009}. 
The star $\beta$ Cephei is the prototype of the $\beta$ Cephei class of
pulsating stars \citep{frost:1903}. Its multi-periodic
photometric and spectroscopic line-profile variability have been studied
extensively \citep{heynderickx:1994, telting:1997,
shibahashi:2000}. In addition to the main pulsation
period of 4h 34m, the star exhibits a very significant
period of 12 d in the equivalent width of the ultraviolet resonance lines.
At the time of its discovery by \cite{fischel:1972} with
the OAO-2 satellite, which was confirmed and further investigated by \cite{panek:1976}, the period had not been precisely determined (6 or 12 days).
Later investigations with IUE data \citep{henrichs:1993, henrichs:1998}
left no doubt that the two minima in equivalent width of the
\ion{C}{iv} stellar wind lines, which are separated by 6 d, are unequal,
and that the real period is 12 d.  \cite{henrichs:1993} proposed
that the UV periodicity arises from the 12-day rotational period of the star
and suggested that the stellar wind is modulated by an oblique dipolar
magnetic field at the surface.  A rotational period of
12 days corresponds well with an adopted radius between six and ten solar
radii, given the reported values of 20 -- 27 km~s$^{-1}$ for $v$sin$i$ 
\citep{abt:2002, telting:1997}.

Support for this hypothesis was provided by the striking similarity between
the UV-line behaviour of $\beta$~Cep and of well-known, chemically peculiar
magnetic B stars, such as the B2 V helium-strong star HD~184927
 \citep{barker:1982, wade:1997}. See \cite{henrichs:2012} for a comparison.  A reported (but not
confirmed) average magnetic field strength of $B$ = (810 $\pm$ 170) G for
$\beta$~Cep itself was published by \cite{rudy:1978}, which is  based on data between 1975 and 1976.
This was however a marginal detection with only one point above the 3$\sigma$ level.
It is unlikely that this result was affected by contaminating emission, because the H$\gamma$ line was mostly used, which was never observed in emission. According to the compilation of \cite{panko:1997}, the system was not likely in an emission state during these years.

\begin{figure}[!ht]
\begin{center}
\epsfig{file=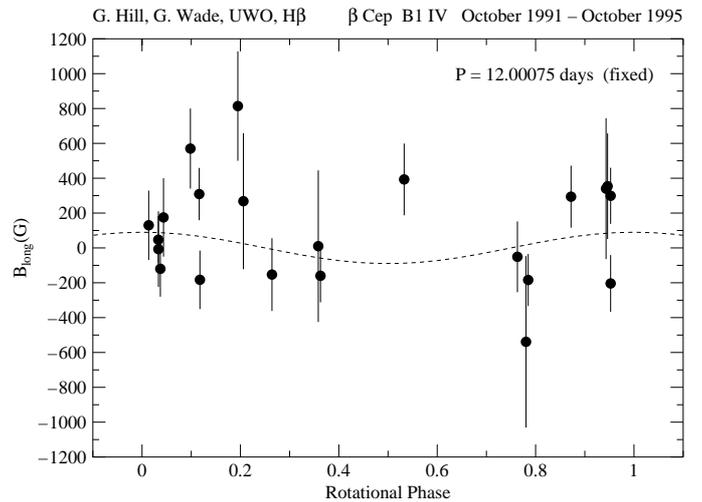,width=\columnwidth,clip=}
\caption{Early magnetic measurements of $\beta$~Cep, which are obtained with the
University of Western Ontario photoelectric Pockels cell polarimeter and
1.2m telescope.  Typical exposure times were between 1 and 3.5 hours.
The dashed curve is the best sinusoid fit, based on the analysis of all data points in this paper, and is the same as in Fig.~\ref{fig:ewuvb}.}
\label{fig:hill} 
\end{center}
\end{figure}

To verify the magnetic hypothesis, \cite{henrichs:1993} presented new
magnetic field measurements which were obtained by one of us (GH), with the University of Western Ontario
photoelectric Pockels cell polarimeter and 1.2m telescope that were simultaneously
measured with UV spectroscopy with the IUE satellite.  The technique used to measure the magnetic
field was differential circular polarimetry in the H$\beta$ line 
(\citealt{landstreet:1982} and references therein).
The 12-day UV period in the equivalent width of the
stellar wind lines of \ion{C}{iv}, \ion{Si}{iii}, \ion{Si}{iv}, and \ion{N}
{v} was confirmed, but the values for the magnetic field
with 1$\sigma$ error bars of about 150 G which are comparable to the measured field
strength, were much lower than the value reported by \cite{rudy:1978}.
However, additional magnetic measurements with the same instrumentation by
G.~Hill and G.~Wade (see Fig.~\ref{fig:hill}) could not confirm the 12-day period which
was unexplained.  It remained puzzling why these new magnetic field
measurements indicated a much lower field than in 1978.  It was suggested 
that the new Be phase of the star, which was discovered in July
1990 by \cite{mathias:1991}, \citep[see also][]{kaper:1992, kaper:1995}, might have been related to the
decrease in magnetic field strength. This hypothesis could not be tested and has since been made obsolete 
because \cite{schnerr:2006e} discovered that the emission stems from
the binary companion. One possible explanation of the discordance between the magnetic field
measurements from the H$\beta$ line and those given below is that H$\beta$ may have been partially filled in with emission during the time of the observations, because H$\alpha$ was in emission at that time.

These considerations motivated us to undertake new magnetic measurements of
$\beta$~Cep with the much more sensitive MuSiCoS polarimeter at the Pic du
Midi observatory in France. Using this instrument has the clear advantage
that all available (mostly metallic) magnetically sensitive lines can be selected in the spectrum,
instead of just one Balmer line, which may be contaminated with some
emission. This paper presents and discusses the observed measurements and 
analyses the correlation between the magnetic, pulsational, and UV behaviour.

The magnetic field of $\beta$ Cep was first measured on December 13, 1998 and was
followed by 22 observations in January 1999 and June/July 1999, such that 
the rotational period was sufficiently covered to allow a thorough assessment.
In an earlier stage of writing the current paper, we decided to separately publish the model
calculations based on these first 23 measurements \citep{donati:2001}, which would
take much less time to complete
in view of the extended data analysis and difficult interpretation of the odd behaviour of the
variable H$\alpha$ emission to follow.
In that paper, we also included a discussion of the
stellar parameters, which we have summarized in Table \ref{param},  except for the Hipparcos distance that was revised by \cite{vanleeuwen:2007}. The model calculations
were constrained by the observed X-ray emission, as observed by \cite{berghofer:1996}.
 \cite{donati:2001} also devoted a discussion to
possible implications of the magnetic field of $\beta$ Cep for
understanding the Be phenomenon, which now appears to be 
academic for this star.
As the variability of the H$\alpha$ line stems from the companion of $\beta$ Cep,
we do not discuss this aspect in the current paper (For a summary, see \citealt{wheelwright:2009}.).
\begin{table}
\begin{center}
\caption[Stellar parameters]{Adopted stellar parameters for \object{$\beta$ Cep}.  The revised distance is taken from \cite{vanleeuwen:2007}.}
\label{param}
\begin{tabular}{ll}
\hline
\hline
Spectral Type            &  B1 IV             \\
$V$                      & 3.2 $\pm$ 0.1      \\
$d_{\rm Hipparcos}$ (pc) & 210 $\pm$ 13       \\
$M_{V}$                  & $-$5.8 $\pm$ 0.2   \\
$M_{\rm bol}$            & $-$5.57 $\pm$ 0.49 \\
log($L$/$L_{\odot}$)     & 4.12 $\pm$ 0.20    \\
$T_{\rm eff}$            & 26 000 K           \\
log$g$ (cm s$^{-2}$)     & 3.7                \\
$R/R_{\odot}$            & 6.5 $\pm$ 1.2      \\
$M/M_{\odot}$            & 12 $\pm$ 1         \\
$v$sin$i$ (km\,s$^{-1}$) & 27 $\pm$ 4         \\
$P_{\rm rot}$ (d)        & 12.000752 $\pm$ 0.000107 \\
\hline
\hline
\end{tabular}
\end{center}
\end{table}

In Sect.~\ref{observations}, we describe the experimental setup and the observations.
Sect.~\ref{data} summarizes the data reduction and the results. In Sect.~\ref{period},
we derive the system velocity and compare
the known periodicities in $\beta$~Cep with the magnetic measurements.
In the last section, we give our
conclusions and discuss the implications of the current measurements, including a discussion of recent interferometric and X-ray measurements.

\section{Experimental setup and observations}
\label{observations}

We obtained circular polarisation (Stokes $V$) and total intensity (Stokes $I$) of
$\beta$~Cep, using the MuSiCoS spectropolarimeter mounted on the 2-m
T\'elescope Bernard Lyot (TBL) at the Observatoire du Pic du Midi. The strategy of the
observations was fourfold: (1) to cover the known 12-day period of the UV
lines, (2) to have reasonable coverage during one pulsational period of
4.6~h, (3) to study the magnetic and pulsational behaviour on a yearly timescale, and
(4) to investigate the systemic radial velocity of the star in its 90-year orbit.  The journal
of observations is given in Table~\ref{tab:results}.

The spectropolarimetric setup consisted of the MuSiCoS fiber-fed cross
dispersed \'echelle spectrograph \citep{baudrand:1992, catala:1993}
with a dedicated polarimetric unit \citep[described by][]{donati:1999}
mounted at the Cassegrain focus.
The light passes through a rotatable quarter wave plate, converting the circular polarisation into
linear, after which the beam is split into two beams with a linear polarisation along and
perpendicular to the instrumental reference azimuth.
Two fibers transport the light to the spectrograph, where
both orthogonal polarisation states are simultaneously recorded. The
spectral coverage in one exposure is from 450 to 660 nm with a resolving
power of about 35000. The Site CCD detector with 1024$\times$1024 of
24$\mu$ pixels was used, which has a quantum efficiency exceeding 50\% in
the U band.

A complete Stokes $V$ measurement consists of four subsequent subexposures
between which the quarter wave plate is rotated, so that a sequence is obtained
with $-$45/45/45/$-$45 degrees angle (called the $q1$-$q3$-$q3$-$q1$ sequence). This results 
in exchange of the two beams throughout the whole instrument. With this sequence, all systematic spurious
circular polarisation signals down to 0.002$\%$ rms can be suppressed
\citep{donati:1997, donati:1999, wade:2000}. At the beginning
and at the end of each night, we took 15 flatfield exposures, whereas
the usual bias frames and polarisation check exposures, and wavelength calibrations with a Th-Ar lamp were taken several times per night.  For
the reduction, we used the flatfield series averaged over the night, where we separated the flatfields taken in the $q1$ and $q3$ orientation.

\section{Data reduction and results}
\label{data}

The data reduction was done with the dedicated ESpRIT reduction package,
described by \cite{donati:1997}. We used a slightly modified version in which the separated $q1$ and $q3$ flatfields are used.  With this package, the geometry of
the orders on the CCD is first determined, and after an automatic
wavelength calibration on the Th-Ar frames, a rigorous optimal extraction
of the orders is performed.
The method we used to calculate the magnetic field strength includes a least-squares deconvolution (LSD) to calculate a normalised average Stokes $I$ line profile and corresponding Stokes $V$ line profiles. This magnetic field strength calculation uses as many possible spectral lines. The presence of a magnetic field will result in a typical Zeeman signature in the average Stokes $V$ profile, from which the effective longitudinal component ($B_{\ell}$) of the stellar magnetic field can be determined. This is done by taking the first-order moment using the well-known relation
\citep{mathys:1989, donati:1997},

\begin{equation}
\label{equ:bl}
B_{\ell} = (-2.14\times 10^{11}{\rm G})\frac{\int vV(v) dv}
{\lambda g c\int [1 - I(v)] dv},
\end{equation}

where $\lambda$, in nm, is the mean wavelength, $c$ is the velocity of
light in the same units as the velocity $v$, and $g$ is the mean value of the Land\'e
factors of all lines used to construct the LSD profile. We used $\lambda$ =
512.5 nm and $g$ = 1.234.
The noise in the LSD spectra was measured and given in
Table~\ref{tab:results}, along with the signal-to-noise ratio obtained in
the raw data.  The profiles were normalised outside the regions of [$-$120,
120] km~s$^{-1}$.

Considerations of several important effects appeared decisive for the
final outcome of the magnetic results. The absolute values of $B_{\ell}$ are impacted
by: fringe patterns, which are present in many spectra;
the selection of spectral lines for the LSD analysis; the limits of integration
in Eq.~\ref{equ:bl}; and the asymmetry of the lines thanks to the pulsation, which we discuss in turn.

\subsection{Correction for fringes}
\begin{figure}[t!]
\begin{center}
\epsfig{file=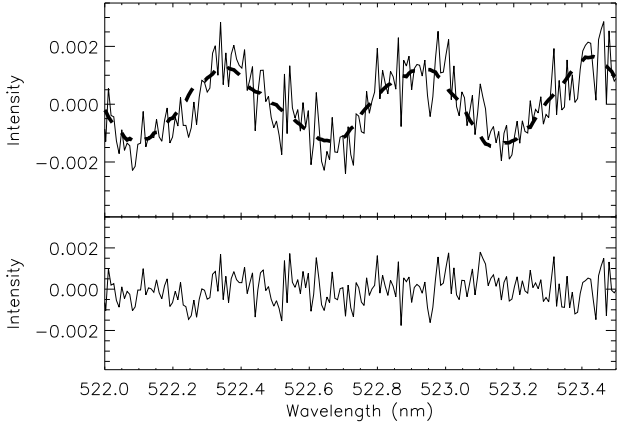,width=0.9\columnwidth,clip=}
\caption{{\it Top:} Overplot of a smoothed Stokes $V$ spectrum of Vega
(dashed line) and a Stokes $V$ spectrum of $\beta$ Cep (1999, nr. 22).
Both contain a strong spurious modulation caused by fringing on the CCD.
{\it Bottom:} Stokes $V$ spectrum of $\beta$ Cep after removal of the fringes.}
\label{fig:fringes}
\end{center}
\end{figure}

Many Stokes $V$ spectra appeared to be strongly
affected by interference fringes that are to likely originate in the quarter-wave plate.
These fringes could
induce a spurious Zeeman detection that could modify the value of the
measured longitudinal magnetic field.  We have eliminated the fringes from
the spectra using a Stokes $V$ spectrum of a bright star that carried no detectable 
Zeeman signature. We used spectra of Vega during the same observing 
run for all but two datasets. For the 2001 data, we used $\zeta$ Oph.
We could not apply the fringe correction to the dataset obtained in June 2000, because no suitable
spectrum of Vega or a similar star was available.  This template
spectrum is smoothed with a running mean of 10 or 20 points to remove any
possible features that could modify the $\beta$ Cep Stokes $V$ spectra
during the correction process.  In this way, we assure that only the
(sine-like wave) fringe pattern remains.  The spectra to be corrected are then
divided by the template. The method is illustrated in
Fig.~\ref{fig:fringes} showing an overplot of the smoothed Vega spectrum
and the Stokes $V$ spectrum nr.\ 22 (1999) of $\beta$ Cep, with the
resulting spectrum.  
After application of this procedure, the error bars improved by typically 
10\% to 20\%, whereas the resulting magnetic values shifted dramatically by
more than 50 G in some cases, which showed the necessity to correct for the fringes.
Table \ref{bfieldvsvrot} gives numerical examples, including the cumulative effects of the
selected line list, as discussed below. We note that \cite{wade:2006} used a wavelet transform
procedure to remove the fringe patterns by using the same spectrum of Vega in 2003 to analyse the
spectra of $\alpha$ And.

\onltab{
\onecolumn
\begin{longtab}
\begin{longtable}{llccccccrrrcrc}
\caption[]{Observations and results of magnetic measurements from
MuSiCoS spectropolarimetry of $\beta$~Cep at TBL at the Pic du Midi, 1998 -- 2005.
The first column identifies the sequence number of a full set of four 
subexposures in the given observing period. In case of missing numbers (in 2001), the obtained
intermediate spectra (less than four) were used for the radial velocity study.
The Barycentric Julian date is given at the center of the total exposure time
($t_{\rm exp}$). 
Column 5 lists the quality of the Stokes $V$ spectra, which is
expressed as the S/N per 4.5 km s$^{-1}$ around 550 nm in the raw spectrum,
and the relative rms noise level N$_{\rm LSD}$ (per 4.5 km s$^{-1}$
velocity bin) in the Least-Squares Deconvolved Stokes $V$ spectra.
The phase in the radial velocity curve (with phase 0 defined at
maximum) has been calculated with the ephemeris given by \cite{pigulski:1992}
 in column 7. The UV (rotational) phase in column 8
has been derived from Eq.~\ref{eq:uvphase}. The measured radial velocity
(accuracy: 2.5 km s$^{-1}$) is given in column 9, whereas the velocity
shift, which is measured at minimum flux and used before calculating the magnetic field,
is given in column 10. Columns 11 and 12 give the magnetic field values
with their 1$\sigma$ uncertainties. The last two columns give the computed magnetic values
of the diagnostic null (or $N$) spectrum with their 1$\sigma$ uncertainties.}\\
\hline
\hline
Nr. & \multicolumn{1}{c}{Date} &HJD & $t_{\rm exp}$ & \multicolumn{1}{c}{S/N} &
   N$_{\rm LSD}$ & Puls. & UV&\multicolumn{1}{c}{$v_{\rm rad}$} &
\multicolumn{1}{c}{$v_{\rm min}$} & \multicolumn{1}{c}{$B_{l}$} & $\sigma(B_{l})$
   & \multicolumn{1}{c}{$N_{l}$} & $\sigma(N_{l})$\\
   & &$-$2451100 &  min      & \multicolumn{1}{c}{pxl$^{-1}$} &$\%$& Phase& Phase&\multicolumn{1}{c}{km\,s$^{-1}$}
   &\multicolumn{1}{c}{km\,s$^{-1}$}& \multicolumn{1}{c}{G} & G
   & \multicolumn{1}{c}{G} & \multicolumn{1}{c}{G}\\
\hline
\endfirsthead
\caption{Continued.} \\
\hline 
Nr. & \multicolumn{1}{c}{Date} &HJD & $t_{\rm exp}$ & \multicolumn{1}{c}{S/N} &
   N$_{\rm LSD}$ & Puls. & UV&\multicolumn{1}{c}{$v_{\rm rad}$} &
\multicolumn{1}{c}{$v_{\rm min}$} & \multicolumn{1}{c}{$B_{l}$} & $\sigma(B_{l})$
   & \multicolumn{1}{c}{$N_{l}$} & $\sigma(N_{l})$\\
   & &$-$2451100 &  min      & \multicolumn{1}{c}{pxl$^{-1}$} &$\%$& Phase& Phase&\multicolumn{1}{c}{km\,s$^{-1}$}
   &\multicolumn{1}{c}{km\,s$^{-1}$}& \multicolumn{1}{c}{G} & G
   & \multicolumn{1}{c}{G} & \multicolumn{1}{c}{G}\\ 
\hline
\endhead
\hline
\endfoot 
\01 & 1998 Dec. 13 & \061.340 & 20 &  290 & 0.057 & 0.092 & 0.600 & $-$23.8 & $-$24.2 & $-$98 & 55 &    37 & 55\\
\02 & 1998 Dec. 14 & \062.337 & 40 &  310 & 0.052 & 0.331 & 0.683 &  $-$4.9 &  $-$0.6 & $-$56 & 51 &    22 & 51\\
\03 & 1998 Dec. 15 & \063.335 & 40 & 1160 & 0.013 & 0.567 & 0.766 & $-$14.1 & $-$13.2 &    33 & 13 &    11 & 13\\
\04 & 1998 Dec. 16 & \064.345 & 40 &  450 & 0.034 & 0.870 & 0.851 & $-$34.5 & $-$38.9 &    71 & 33 &     3 & 33\\
\05 & 1998 Dec. 17 & \065.246 & 20 &  920 & 0.015 & 0.601 & 0.926 & $-$16.3 & $-$16.3 &    81 & 15 &     2 & 15\\
\06 & 1998 Dec. 17 & \065.342 & 30 & 1050 & 0.013 & 0.103 & 0.934 & $-$23.0 & $-$24.4 &    85 & 12 & $-$16 & 12\\
\07 & 1998 Dec. 18 & \066.256 & 40 &  740 & 0.021 & 0.904 & 0.010 & $-$35.0 & $-$39.7 &    66 & 20 & $-$29 & 20\\
\08 & 1998 Dec. 18 & \066.290 & 30 &  770 & 0.020 & 0.080 & 0.013 & $-$23.4 & $-$25.0 &    64 & 19 & $-$14 & 19\\
\09 & 1998 Dec. 18 & \066.314 & 30 &  950 & 0.015 & 0.207 & 0.015 & $-$12.3 &  $-$9.8 &    91 & 14 &     0 & 13\\
 10 & 1998 Dec. 18 & \066.338 & 30 &  940 & 0.013 & 0.334 & 0.017 &  $-$3.4 &     1.1 &   117 & 13 &  $-$9 & 13\\
 11 & 1998 Dec. 18 & \066.368 & 27 &  910 & 0.015 & 0.489 & 0.019 &  $-$7.1 &  $-$3.3 &   114 & 14 &     6 & 14\\
 12 & 1999 Jan. 13 & \092.256 & 40 &  760 & 0.020 & 0.397 & 0.176 &  $-$4.7 &     0.2 &    61 & 20 &     6 & 19\\
 13 & 1999 Jan. 15 & \094.256 & 20 &  500 & 0.030 & 0.893 & 0.343 & $-$34.9 & $-$40.5 & $-$50 & 29 &     9 & 29\\
 14 & 1999 Jan. 24 &  103.263 & 20 &  660 & 0.023 & 0.180 & 0.094 & $-$13.8 & $-$12.2 &   116 & 22 & $-$46 & 22\\
 15 & 1999 Jan. 25 &  104.271 & 24 &  530 & 0.027 & 0.474 & 0.178 &  $-$5.5 &  $-$0.9 &    81 & 26 &     0 & 27\\
 16 & 1999 June 30 &  259.504 & 40 &  690 & 0.022 & 0.406 & 0.113 &  $-$2.6 &     1.8 &    63 & 22 &  $-$5 & 22\\
 17 & 1999 June 30 &  259.545 & 60 &  640 & 0.022 & 0.622 & 0.116 & $-$21.5 & $-$21.8 &    55 & 21 &    22 & 21\\
 18 & 1999 June 30 &  260.492 & 40 &  670 & 0.021 & 0.590 & 0.195 & $-$17.6 & $-$16.2 &    73 & 21 &     6 & 21\\
 19 & 1999 July 1  &  260.527 & 50 &  770 & 0.019 & 0.774 & 0.198 & $-$31.3 & $-$33.6 &    21 & 18 & $-$23 & 18\\
 20 & 1999 July 3  &  262.512 & 60 &  830 & 0.018 & 0.197 & 0.363 &  $-$8.4 &  $-$8.6 & $-$47 & 17 &    11 & 17\\
 21 & 1999 July 3  &  263.486 & 60 &  680 & 0.019 & 0.312 & 0.445 &  $-$1.4 &     2.0 & $-$31 & 18 & $-$27 & 17\\
 22 & 1999 July 6  &  266.467 & 40 &  760 & 0.014 & 0.960 & 0.693 & $-$28.1 & $-$29.7 & $-$22 & 14 & $-$30 & 13\\
 23 & 1999 July 7  &  267.417 & 60 &  790 & 0.012 & 0.950 & 0.772 & $-$27.7 & $-$30.7 &  $-$2 & 11 & $-$26 & 11\\

\hline
\01 & 2000 June 17 &  612.632 & 40 & 890 & 0.014 & 0.241 & 0.538 &     7.9 &     3.2 & $-$69 & 20 &  $-$2 & 17\\
\02 & 2000 June 21 &  616.592 & 48 & 930 & 0.014 & 0.030 & 0.868 & $-$20.7 & $-$16.7 &    94 & 21 &    27 & 19\\
\03 & 2000 June 26 &  621.569 & 40 & 980 & 0.013 & 0.158 & 0.283 &  $-$0.7 &  $-$3.4 &    31 & 18 & $-$18 & 16\\
\04 & 2000 June 26 &  621.600 & 40 & 980 & 0.013 & 0.320 & 0.286 &     7.0 &     3.6 &    19 & 19 &     8 & 15\\
\05 & 2000 June 28 &  623.641 & 28 & 640 & 0.021 & 0.035 & 0.456 & $-$18.0 & $-$15.5 &  $-$4 & 31 &  $-$7 & 29\\
\06 & 2000 June 29 &  624.635 & 40 & 910 & 0.014 & 0.253 & 0.539 &     4.4 &     0.3 & $-$35 & 21 &    15 & 18\\
\07 & 2000 June 30 &  625.643 & 40 & 800 & 0.017 & 0.545 & 0.623 &  $-$8.6 & $-$10.2 & $-$59 & 26 & $-$23 & 21\\
\08 & 2000 July  5 &  630.614 & 20 & 530 & 0.025 & 0.642 & 0.037 & $-$21.9 & $-$19.8 &   137 & 37 & $-$10 & 34\\
\09 & 2000 July  5 &  630.631 & 20 & 540 & 0.025 & 0.731 & 0.038 & $-$30.0 & $-$26.0 &    84 & 38 &    54 & 36\\
 10 & 2000 July  5 &  630.648 & 20 & 500 & 0.028 & 0.820 & 0.040 & $-$33.1 & $-$28.3 &    16 & 40 & $-$16 & 38\\
 11 & 2000 July  6 &  631.615 & 20 & 490 & 0.028 & 0.897 & 0.120 & $-$29.7 & $-$26.7 &    70 & 40 &     1 & 38\\
 12 & 2000 July  6 &  631.651 & 20 & 380 & 0.035 & 0.086 & 0.123 &  $-$9.9 & $-$10.6 &   107 & 50 & $-$60 & 49\\
 13 & 2000 July  7 &  632.667 & 20 & 500 & 0.026 & 0.419 & 0.208 &     6.6 &     1.7 &   100 & 38 &     6 & 37\\
 14 & 2000 July  8 &  634.464 & 28 & 680 & 0.018 & 0.853 & 0.358 & $-$34.2 & $-$28.2 & $-$41 & 26 &    11 & 24\\
 15 & 2000 July 13 &  638.621 & 20 & 450 & 0.029 & 0.677 & 0.704 & $-$23.5 & $-$22.4 & $-$130& 46 &     11& 43\\
 16 & 2000 July 13 &  638.638 & 20 & 460 & 0.028 & 0.766 & 0.705 & $-$30.1 & $-$27.9 & $-$20 & 43 & $-$11 & 39\\
 17 & 2000 July 17 &  642.581 & 20 & 600 & 0.021 & 0.465 & 0.034 &  $-$2.7 &  $-$5.1 &    93 & 31 & $-$35 & 30\\
 18 & 2000 July 17 &  642.598 & 20 & 670 & 0.020 & 0.555 & 0.035 & $-$13.3 & $-$13.4 &    66 & 30 &     8 & 28\\
 19 & 2000 July 17 &  642.615 & 20 & 630 & 0.021 & 0.644 & 0.037 & $-$23.6 & $-$21.3 &   136 & 31 &     9 & 30\\
 20 & 2000 July 17 &  642.631 & 20 & 580 & 0.023 & 0.728 & 0.038 & $-$31.3 & $-$27.3 &   164 & 34 &     8 & 33\\
 21 & 2000 July 17 &  642.648 & 20 & 630 & 0.020 & 0.817 & 0.040 & $-$32.2 & $-$27.9 &    94 & 30 & $-$26 & 29\\
\hline

\01 & 2001 June 19 &  979.628 & 20 & 560 & 0.028 & 0.878 & 0.119 &     $-$25.5    &$-$29.8         &  $-$62 & 28 &    12 & 26\\
\02 & 2001 June 19 &  979.644 & 20 & 630 & 0.025 & 0.963 & 0.121 &     $-$20.5    &$-$21.9         &  $-$35 & 24 & $-$33 & 22\\
\03 & 2001 June 20 &  980.602 & 20 & 580 & 0.027 & 0.991 & 0.201 & $-$14.1 & $-$12.9 &     27 & 26 & $-$15 & 25\\
\04 & 2001 June 20 &  980.619 & 20 & 590 & 0.026 & 0.080 & 0.202 &  $-$5.2 &  $-$1.8 &      8 & 26 &     4 & 25\\
\05 & 2001 June 20 &  980.637 & 20 & 590 & 0.026 & 0.175 & 0.203 &     2.8 &     7.7 &  $-$15 & 25 &    38 & 25\\
\06 & 2001 June 21 &  981.612 & 20 & 730 & 0.020 & 0.293 & 0.285 &     7.1 &    10.9 &  $-$57 & 19 &     4 & 18\\
\07 & 2001 June 21 &  981.630 & 20 & 600 & 0.024 & 0.388 & 0.286 &     3.6 &     6.5 &  $-$16 & 23 &  $-$2 & 23\\
\08 & 2001 June 21 &  981.656 & 16 & 480 & 0.034 & 0.524 & 0.288 &  $-$2.2 &  $-$0.3 &      0 & 34 &    10 & 34\\
\09 & 2001 June 22 &  982.626 & 20 & 740 & 0.021 & 0.617 & 0.369 & $-$15.9 & $-$20.1 &  $-$96 & 21 &    13 & 20\\
 10 & 2001 June 22 &  982.643 & 20 & 660 & 0.023 & 0.706 & 0.371 & $-$22.2 & $-$27.8 & $-$156 & 23 &    36 & 22\\
 11 & 2001 June 23 &  983.628 & 20 & 240 & 0.073 & 0.877 & 0.453 & $-$24.6 & $-$30.3 & $-$196 & 73 &   102 & 72\\
 13 & 2001 June 24 &  984.554 & 20 & 650 & 0.024 & 0.739 & 0.530 & $-$25.7 & $-$29.6 & $-$186 & 24 & $-$25 & 23\\
 14 & 2001 June 24 &  984.571 & 20 & 710 & 0.022 & 0.828 & 0.531 & $-$25.9 & $-$30.2 & $-$166 & 22 &    11 & 21\\
 15 & 2001 June 25 &  985.523 & 20 & 610 & 0.027 & 0.826 & 0.611 & $-$26.1 & $-$29.9 & $-$140 & 27 &    14 & 25\\
 16 & 2001 June 25 &  985.540 & 20 & 630 & 0.026 & 0.915 & 0.612 & $-$21.3 & $-$23.4 & $-$138 & 25 & $-$24 & 23\\
 17 & 2001 June 26 &  986.541 & 20 & 600 & 0.027 & 0.170 & 0.695 &     3.2 &     8.5 &  $-$60 & 27 &    51 & 25\\
 19 & 2001 June 26 &  986.562 & 20 & 300 & 0.055 & 0.280 & 0.697 &     9.2 &    15.1 & $-$104 & 55 &    88 & 54\\
 20 & 2001 June 27 &  987.485 & 20 & 270 & 0.062 & 0.126 & 0.774 &     0.5 &     2.8 &      3 & 61 &    15 & 61\\
 21 & 2001 June 27 &  987.502 & 20 & 410 & 0.040 & 0.215 & 0.776 &     5.8 &     9.1 &     35 & 39 & $-$39 & 38\\
 22 & 2001 June 27 &  987.520 & 20 & 400 & 0.040 & 0.310 & 0.777 &     7.7 &    11.5 &  $-$15 & 40 &  $-$1 & 41\\
 23 & 2001 June 29 &  989.531 & 20 & 570 & 0.027 & 0.867 & 0.945 & $-$23.7 & $-$28.4 &     91 & 27 & $-$21 & 26\\
 24 & 2001 June 29 &  989.548 & 20 & 540 & 0.030 & 0.956 & 0.946 & $-$17.4 & $-$21.0 &     68 & 30 &    15 & 29\\
 25 & 2001 June 30 &  990.533 & 20 & 690 & 0.023 & 0.128 & 0.028 &  $-$2.2 &  $-$0.8 &     57 & 23 & $-$25 & 22\\
 26 & 2001 June 30 &  990.553 & 20 & 670 & 0.022 & 0.233 & 0.030 &     5.5 &     9.4 &     67 & 22 & $-$26 & 22\\
 27 & 2001 July  1 &  991.573 & 20 & 560 & 0.027 & 0.588 & 0.115 & $-$13.4 & $-$13.7 &     51 & 26 & $-$21 & 26\\
 28 & 2001 July  1 &  991.591 & 20 & 640 & 0.024 & 0.682 & 0.116 & $-$22.1 & $-$24.3 &    100 & 24 &    34 & 24\\
 29 & 2001 July  2 &  992.535 & 20 & 620 & 0.026 & 0.638 & 0.195 & $-$20.0 & $-$21.4 &     12 & 25 & $-$31 & 25\\
 30 & 2001 July  2 &  992.554 & 20 & 560 & 0.030 & 0.738 & 0.197 & $-$25.9 & $-$29.0 &     73 & 29 & $-$44 & 28\\
 31 & 2001 July  3 &  993.539 & 20 & 730 & 0.023 & 0.909 & 0.279 & $-$21.1 & $-$24.4 &    -36 & 22 &  $-$2 & 21\\
 32 & 2001 July  3 &  993.557 & 20 & 750 & 0.022 & 0.003 & 0.280 & $-$13.1 & $-$14.2 &    -44 & 21 &    40 & 20\\
   \hline											
\01 & 2002 June 11 & 1336.517 & 20 & 570 & 0.027 & 0.457 & 0.858 &     3.4 &     5.3 &     99 & 27 & $-$34 & 27\\
\02 & 2002 June 12 & 1337.521 & 20 & 520 & 0.032 & 0.728 & 0.942 & $-$23.3 & $-$28.4 &     46 & 33 &    56 & 32\\
\03 & 2002 June 13 & 1338.537 & 28 & 450 & 0.035 & 0.062 & 0.027 &  $-$6.6 &  $-$7.3 &     88 & 34 & $-$12 & 34\\
\04 & 2002 June 14 & 1339.547 & 28 & 530 & 0.031 & 0.364 & 0.111 &     7.7 &    12.5 &     86 & 31 &    28 & 30\\
\05 & 2002 June 15 & 1340.511 & 20 & 400 & 0.039 & 0.425 & 0.191 &     3.6 &     7.8 &     25 & 39 &    64 & 39\\
\06 & 2002 June 16 & 1341.529 & 20 & 540 & 0.031 & 0.769 & 0.276 & $-$24.8 & $-$30.0 &  $-$36 & 30 & $-$33 & 29\\
\07 & 2002 June 17 & 1342.508 & 20 & 510 & 0.030 & 0.908 & 0.358 & $-$21.9 & $-$26.8 &  $-$61 & 29 & $-$19 & 29\\
\08 & 2002 June 17 & 1343.444 & 28 & 710 & 0.022 & 0.822 & 0.436 & $-$24.8 & $-$30.0 &  $-$67 & 22 &  $-$3 & 21\\
\09 & 2002 June 17 & 1343.465 & 28 & 740 & 0.021 & 0.932 & 0.437 & $-$19.8 & $-$23.9 & $-$106 & 21 & $-$29 & 21\\
 10 & 2002 June 18 & 1344.463 & 28 & 570 & 0.026 & 0.172 & 0.520 &     3.1 &     7.0 &  $-$70 & 26 &    24 & 25\\
 11 & 2002 June 18 & 1344.484 & 28 & 700 & 0.023 & 0.282 & 0.522 &     9.9 &    15.9 & $-$120 & 23 &  $-$8 & 23\\
 12 & 2002 June 21 & 1346.524 & 20 & 490 & 0.033 & 0.991 & 0.692 & $-$15.6 & $-$15.1 &  $-$63 & 33 &    65 & 33\\
 13 & 2002 June 21 & 1346.540 & 20 & 510 & 0.031 & 0.075 & 0.694 &  $-$7.4 &  $-$4.2 &  $-$62 & 31 &    34 & 31\\
 14 & 2002 June 22 & 1348.497 & 20 & 690 & 0.023 & 0.349 & 0.857 &     6.2 &     9.5 &     80 & 23 & $-$11 & 23\\
 15 & 2002 June 24 & 1350.438 & 28 & 710 & 0.021 & 0.539 & 0.018 &  $-$8.9 &  $-$7.9 &     44 & 22 & $-$30 & 21\\
 16 & 2002 June 26 & 1351.531 & 28 & 710 & 0.022 & 0.277 & 0.109 &     7.2 &    13.0 &     30 & 23 &     0 & 22\\
 17 & 2002 June 27 & 1352.556 & 28 & 300 & 0.054 & 0.658 & 0.195 & $-$22.3 & $-$24.1 &      9 & 53 &    15 & 52\\
 18 & 2002 June 27 & 1352.578 & 28 & 540 & 0.031 & 0.773 & 0.197 & $-$28.0 & $-$30.0 &     52 & 30 &    18 & 30\\
  \hline
\01 & 2003 June 7  & 1697.613 & 27 & 580 & 0.028 & 0.119 & 0.948 &  $-$0.3 &     0.5 &     80 & 27 &    23 & 26\\
\02 & 2003 June 9  & 1700.499 & 27 & 440 & 0.034 & 0.270 & 0.188 &    10.9 &    16.0 &     42 & 34 &    29 & 33\\
\03 & 2003 June 12 & 1702.545 & 28 & 430 & 0.040 & 0.014 & 0.359 & $-$12.6 & $-$15.7 &  $-$73 & 39 &    29 & 37\\
\04 & 2003 June 14 & 1704.575 & 20 & 550 & 0.028 & 0.671 & 0.528 & $-$18.0 & $-$20.3 &  $-$80 & 28 & $-$37 & 28\\
\05 & 2003 June 17 & 1707.647 & 20 & 590 & 0.026 & 0.798 & 0.784 & $-$23.9 & $-$29.3 &     29 & 26 &    12 & 26\\
\06 & 2003 June 18 & 1709.449 & 24 & 640 & 0.023 & 0.258 & 0.934 &     8.5 &    12.7 &    100 & 23 &  $-$9 & 22\\
\07 & 2003 June 18 & 1709.469 & 24 & 730 & 0.021 & 0.361 & 0.936 &     8.9 &    13.4 &     96 & 21 &    11 & 21\\
\08 & 2003 June 20 & 1711.455 & 32 & 510 & 0.032 & 0.787 & 0.101 & $-$26.1 & $-$29.6 &     70 & 32 & $-$29 & 31\\
\09 & 2003 July  8 & 1728.553 & 28 & 550 & 0.029 & 0.550 & 0.526 & $-$11.4 & $-$11.4 &  $-$56 & 30 &    42 & 29\\
 10 & 2003 July 26 & 1746.557 & 40 & 460 & 0.032 & 0.065 & 0.026 & $-$15.6 & $-$16.9 &     91 & 32 &    20 & 31\\
 11 & 2003 July 30 & 1751.504 & 40 & 730 & 0.020 & 0.033 & 0.438 & $-$17.6 & $-$20.1 & $-$148 & 20 &  $-$1 & 20\\
 12 & 2003 Aug.  6 & 1757.536 & 40 & 830 & 0.017 & 0.700 & 0.941 & $-$32.3 & $-$36.4 &     71 & 17 & $-$22 & 16\\
\hline
\01 & 2004 June  2 & 2058.564 & 47 & 830 & 0.021 & 0.024 & 0.025 & $-$19.6 & $-$16.9 &    100 & 21 &  $-$8 & 19\\
\02 & 2004 June  8 & 2064.510 & 47 & 910 & 0.023 & 0.238 & 0.521 &     7.5 &     2.1 &  $-$84 & 23 &     6 & 18\\
\03 & 2004 June 26 & 2082.637 &	24 & 570 & 0.033 & 0.399 & 0.031 &   $-$33.9&	 $-$29.4&   79 &	33 &	46 & 31\\
\04 & 2004 July 1  & 2087.633 &	47 & 670 & 0.028 & 0.630 & 0.447 &  ...&	... &  $-$94 &	29 & $-$31 & 26\\
\05 & 2004 July 2  & 2088.632 &	47 & 780 & 0.024 & 0.872 & 0.531 &...&	...  & $-$153 &	24 &	19 & 21\\
\06 & 2004 July 12 & 2098.508 & 35 & 600 & 0.031 & 0.721 & 0.354 & $-$23.0 & $-$18.2 &  $-$33 & 30 &  $-$6 & 29\\
\07 & 2004 July 15 & 2101.505 & 35 & 910 & 0.021 & 0.455 & 0.603 &    11.9 &     7.2 & $-$101 & 21 & $-$15 & 19\\
\08 & 2004 July 20 & 2106.571 & 35 & 570 & 0.034 & 0.050 & 0.025 & $-$10.3 &  $-$8.5 &     69 & 34 &    19 & 32\\
\09 & 2004 July 25 & 2111.532 & 35 & 700 & 0.026 & 0.095 & 0.439 &  $-$3.7 &  $-$3.3 & $-$151 & 25 & $-$19 & 23\\
 10 & 2004 July 26 & 2112.575 & 35 & 730 & 0.024 & 0.571 & 0.526 &  $-$5.5 &  $-$5.7 & $-$143 & 25 & $-$25 & 23\\
 11 & 2004 July 27 & 2114.500 & 35 & 870 & 0.021 & 0.673 & 0.686 & $-$14.8 & $-$13.6 &  $-$84 & 21 & $-$25 & 19\\
 12 & 2004 July 30 & 2116.502 & 35 & 750 & 0.028 & 0.183 & 0.853 &     7.9 &     5.1 &     41 & 27 &     2 & 25\\
 13 & 2004 July 30 & 2117.446 & 35 & 790 & 0.023 & 0.138 & 0.932 &     3.7 &     2.0 &    110 & 22 &    12 & 21\\
 14 & 2004 Aug. 7  & 2124.507 & 35 & 310 & 0.067 & 0.210 & 0.520 &    14.3 &    10.0 &  $-$43 & 67 &    45 & 64\\
 15 & 2004 Aug. 10 & 2128.459 & 35 & 640 & 0.027 & 0.954 & 0.849 & $-$21.6 & $-$17.3 &     23 & 26 &  $-$4 & 24\\
 16 & 2004 Aug. 12 & 2130.491 & 35 & 880 & 0.021 & 0.622 & 0.019 & $-$12.4 & $-$10.9 &     42 & 21 &  $-$9 & 19\\
 17 & 2004 Aug. 14 & 2132.457 & 35 & 700 & 0.026 & 0.946 & 0.183 & $-$17.9 & $-$17.1 &     20 & 25 & $-$19 & 24\\
 18 & 2004 Aug. 24 & 2142.409 & 35 & 590 & 0.031 & 0.191 & 0.012 &     9.4 &     6.2 &     16 & 31 &    35 & 30\\
 19 & 2004 Nov. 5  & 2215.277 &	47 &   750  & 0.025      & 0.730 & 0.084 &   $-$8.0  &   $-$5.0  &	    108  &	24  &	11   & 24  \\
 20 & 2004 Nov. 17 & 2227.271 & 47 & 900 & 0.021 & 0.694 & 0.083 & $-$16.2 & $-$13.9 &    102 & 21 &     8 & 19\\
 21 & 2004 Nov. 21 & 2231.258 & 47 & 920 & 0.021 & 0.627 & 0.415 &  $-$8.0 &  $-$5.0 &  $-$86 & 21 &    15 & 19\\
 22 & 2004 Nov. 23 & 2233.272 & 47 & 970 & 0.020 & 0.198 & 0.583 &    14.1 &    10.8 &  $-$47 & 20 &     8 & 18\\
 23 & 2004 Nov. 25 & 2235.302 & 47 & 980 & 0.021 & 0.854 & 0.752 & $-$20.9 & $-$17.3 &   $-$8 & 20 & $-$17 & 17\\
 24 & 2004 Nov. 27 & 2237.289 & 47 & 750 & 0.025 & 0.285 & 0.918 &    21.3 &    17.3 &     93 & 25 & $-$24 & 23\\
\hline
\01 & 2005 July 15 & 2466.629 & 20 & 710 & 0.020 & 0.262 & 0.028 &    12.0 &     8.7 &     86 & 26 & $-$14 & 24\\
\02 & 2005 July 15 & 2466.645 & 20 & 750 & 0.019 & 0.349 & 0.030 &    17.4 &    12.8 &     54 & 25 &    26 & 23\\
\hline
\hline
\label{tab:results}
\end{longtable}
\end{longtab}

\twocolumn
} 

\subsection{Effect of the spectral line list}
The profiles of 125 relatively weak lines, which are selected with a table
appropriate for early B stars, were combined by means of the LSD method, as 
described above in the interval [$-$243, 243] km~s$^{-1}$. Many
of these lines are blends of multiplets, leaving effectively 77 distinct
lines. The selection of lines included in the construction of the LSD profiles
appear to have a systematic influence on the absolute value of the
magnetic field. The removal of blends from the list resulted in a larger absolute signal.
We also investigated the effect of using measured
instead of theoretically calculated line depths. The smallest error bar, which likely
resulted in the most
reliable value for  $B_{\ell}$, is obtained when fringe correction is applied, strong blends
are excluded, and measured line depths are used,
as illustrated in Table \ref{bfieldvsvrot}.

\begin{table}[htbp]
\begin{center}
\caption[Targets.]{Illustrative sample calculations of magnetic-field strength
without and with fringe correction and with different selections
of spectral lines and weights (line depths)
used for the construction of the
LSD Stokes $V$ profile. Integration limits of $v_{\rm limit} =$ 54 km s$^{-1}$ have been used.
Symbols have the same meaning as in Table~\ref{tab:results}, from which the examples in the last three lines in the table below have been taken.}  
\begin{tabular}{l@{\ }c@{\ }rccrc}
\hline
\hline
Spectrum& Fringe	& \multicolumn{1}{c}{$B_{l}$} & $\sigma(B_{l})$&N$_{\rm LSD}$&\multicolumn{1}{c}{$N_{l}$} & $\sigma(N_{l})$\\
 & correction& \multicolumn{1}{c}{G} & \multicolumn{1}{c}{G} &$\%$ & \multicolumn{1}{c}{G} 	& \multicolumn{1}{c}{G}	\\
\hline
\multicolumn{7}{l}{With all available lines with theoretical depths:}\\
2003 $\#$12 & no  &    130 &  21 & 0.014 & $-$13 &  19\\
2004 $\#$1  & no  &     77 &  25 & 0.017 &     1 &  22\\
2004 $\#$2  & no  & $-$110 &  34 & 0.022 &  $-$9 &  21\\
2003 $\#$12 & yes &     95 &  19 & 0.013 & $-$13 &  19\\
2004 $\#$1  & yes &     95 &  24 & 0.016 &     1 &  22\\
2004 $\#$2  & yes & $-$107 &  27 & 0.018 &  $-$9 &  21\\
\hline
\multicolumn{7}{l}{With measured line depths and without all strong blends:}\\
2003 $\#12$ & no  &    145 & 19 & 0.018 & $-$22 & 16 \\
2004 $\#1 $ & no  &     83 & 22 & 0.022 &  $-$7 & 19 \\
2004 $\#2 $ & no  &  $-$83 & 29 & 0.029 &     6 & 18 \\
2003 $\#12$ & yes &     71 & 17 & 0.017 & $-$22 & 16 \\
2004 $\#1 $ & yes &    100 & 21 & 0.021 &  $-$8 & 19 \\
2004 $\#2 $ & yes &  $-$84 & 23 & 0.023 &     6 & 18 \\
\hline
\hline
\end{tabular}
\label{bfieldvsvrot}
\end{center}
\end{table}

\subsection{Limits of integration and asymmetry}

\begin{figure}[!ht]
\begin{center}
\epsfig{file=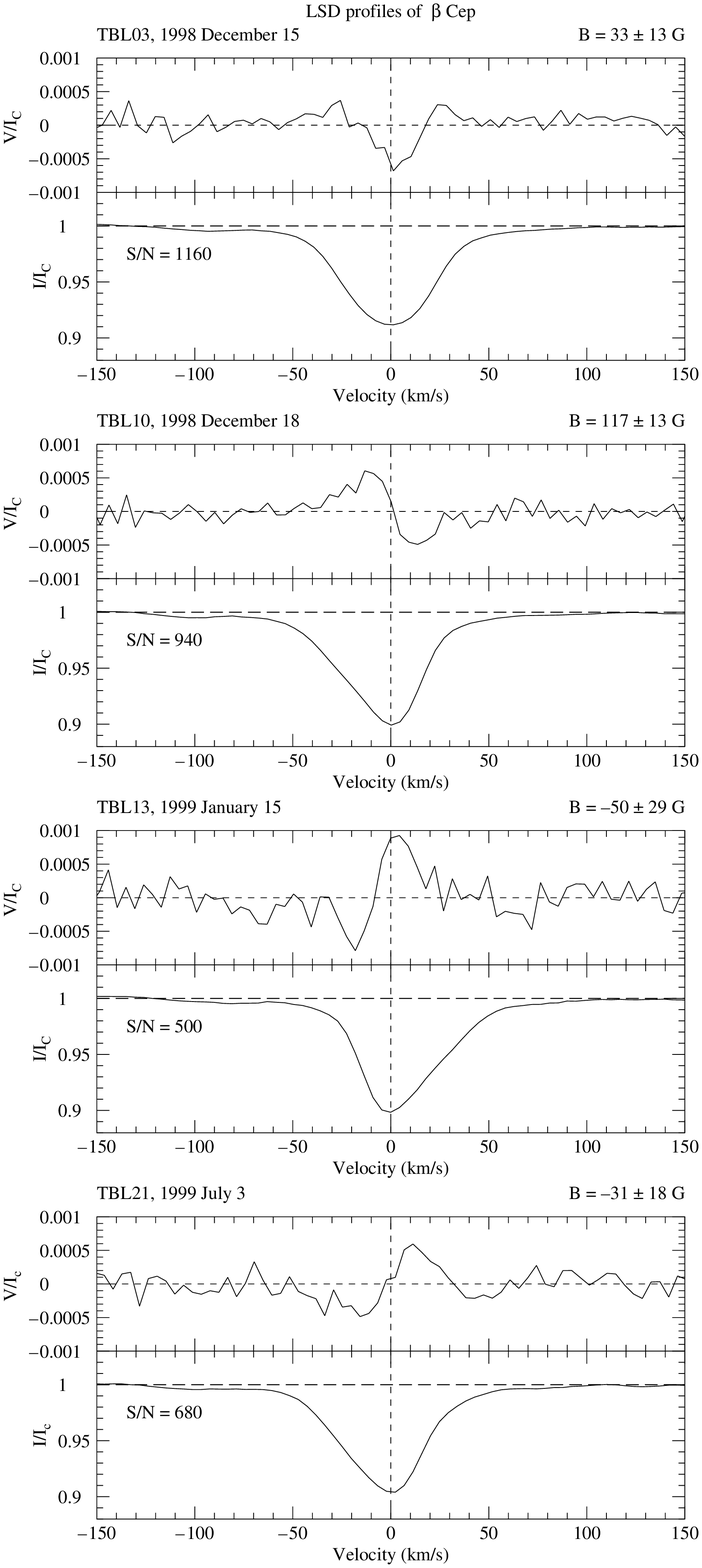,width=0.9\columnwidth,clip=}
\caption{Representative LSD Stokes unpolarised $I$ (lower panel) and
circularly polarised $V$ (upper panel) profiles of $\beta$~Cep from top to
bottom on 15 and 18 December 1998, 15 January, and 3 July 1999. The signal-
to-noise ratio (S/N) per velocity bin in the raw data is indicated.  Note
the clear Zeeman signatures at the zero, positive, and negative fields.  The two lower figures illustrate negative fields at almost
opposite pulsation phases. In Sect.~\ref{period}, we show that the pulsation phase
does not affect the magnetic field determination.
}
\label{fig:lsd}
\end{center}
\end{figure}

Applying Eq. \ref{equ:bl} to calculate $B_{\ell}$ involves taking the first moment, which implies
measuring the asymmetry with respect to the center of the profile. A shift
in the radial velocity scale will therefore affect the value of the magnetic field,
and a proper correction is essential.
Because the radial velocity amplitude is considerable as a consequence of the pulsation of
$\beta$~Cep, we shifted the minima of the $I$ profiles (at
$v_{\rm min}$, which are determined by a parabola fit to the points near the minimum intensity)
to zero velocity before
calculating the longitudinal field strength. The profiles are often
asymmetric, implying that the minimum flux does not occur at the radial
velocity ($v_{\rm rad}$) of the star. The velocity was measured using the first
moment of the profile with respect to the barycentric restframe, which is normalised
by the equivalent width \citep[we followed ][]{schrijvers:1997}. The
measured values of $v_{\rm min}$ and $v_{\rm rad}$ are included in
Table~\ref{tab:results}.

We approximated the integral in Eq.~\ref{equ:bl}
by a simple summation in a range between $\pm v_{\rm limit}$ and computed
the uncertainties as follows:

\begin{equation}
\label{equ:sigbl}
\sigma_{B_\ell} = \left|B_\ell \right| \sqrt{
\frac{\sum {\sigma_{I_i}}^2}{{\left( \sum {1-I_i}\right)}^2}+
\frac{\sum {\sigma_{v_i}}^2 {\sigma_{V_i}}^2}
{{\left( \sum v_i V_i \right)}^2} }.
\end{equation}

The limits of the integral in Eq.~\ref{equ:bl} were carefully determined 
to minimize the uncertainties. We first computed the $B_\ell$ values
for 17 different limits between 10 and 90 km~s$^{-1}$. The full Zeeman
signature is obtained when the maximum value for $B_\ell$ is reached. We
adopted the average optimum value of several test cases, which was at
$v_{\rm limit}$ = 54 km~s$^{-1}$.

\begin{figure*}[ht!]
\begin{center}
\epsfig{file=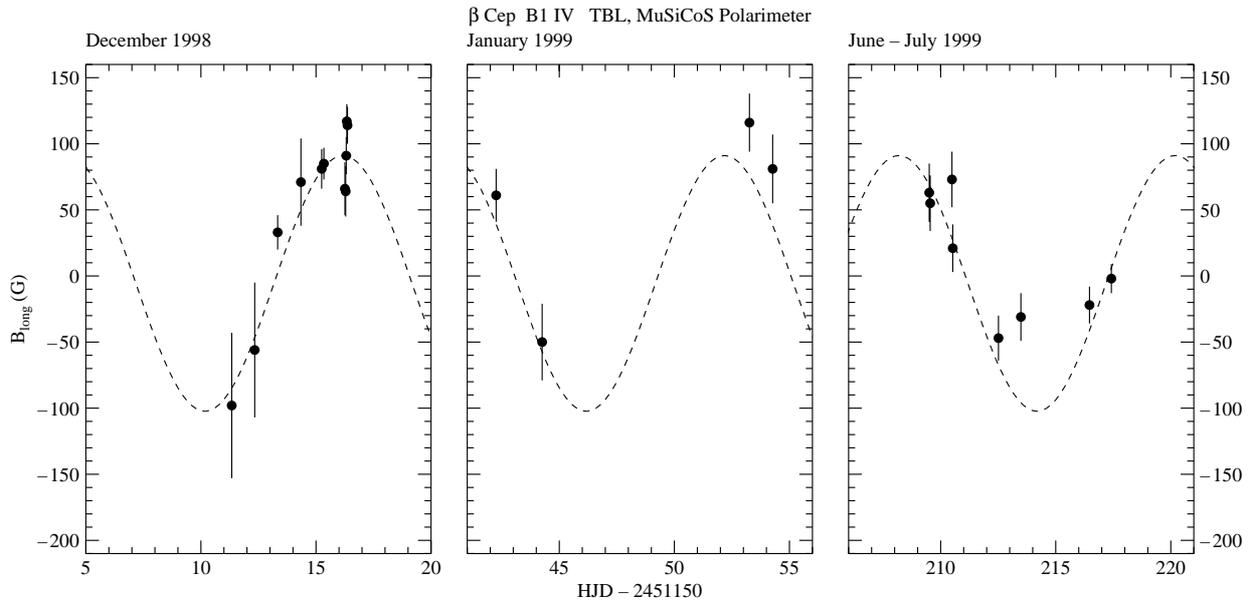,width=1.8\columnwidth,clip=}
\caption{The discovery observations, which include the first 23 datapoints from 
December 1998 until July 1999. 
The longitudinal component of the averaged surface magnetic field
of $\beta$~Cep is plotted. The dashed
curve is the best-fit sine curve for all magnetic data with a fixed
period of 12.00075 d, as derived from UV data.}
\label{fig:bfield}
\end{center}
\end{figure*}

We have also investigated the effect of the asymmetry of the lines (because of
the pulsation) by varying the reference center.  We find that a displacement
of more than $\pm$18 km~s$^{-1}$ gives significant lower values for the
field strength, but the determined values and the
error bars remain constant within 4\% within this range.
For typical examples of Zeeman LSD profiles (i.e.\ the Stokes $I$ and $V$ spectrum
in velocity space), see Fig.~\ref{fig:lsd}. In this figure, we show the LSD profiles for a
zero, positive, and negative field, where the latter are at two
different pulsation phases of the star (see below).

We also calculated diagnostic null (or $N$)
spectra, which are associated with each Stokes $V$ spectrum, by using subexposures
with identical waveplate orientations.  This should provide an accurate
indication of the noise and should not give a detectable signal.
Upon examination of the $N$
profiles, we find that the magnetic fields measured from these spectra are in most cases
consistent with zero, in spite of
some spurious signals, which are not related to the $V$ profiles. The
variable asymmetry in the line profiles has therefore a minor effect on the
magnetic field measurements. The effect of asymmetry was thoroughly investigated 
by \cite{schnerr:2006a} for the pulsating B star \object{$\nu$ Eri}, which showed much 
stronger spurious signals. 

In Table~\ref{tab:results}, we collected the final
results of our calculations, which
included the values calculated from the $N$ spectra.
In Fig.~\ref{fig:bfield}, the discovery observations during the first year are plotted as a function of time, along
with the best-fit sinusoid to all magnetic data (see below).

\section{Period analysis}
\label{period}

The two main periodicities in $\beta$~Cep are 4h 34m of
the radial pulsation mode \citep{frost:1903}, and 12 d 
in the UV wind lines, which has been known since 1972 \citep{fischel:1972}.  We investigate whether the
observed magnetic variability is related to these periods.

\subsection{UV stellar wind period}

\begin{figure}[t!]
\begin{center}
\epsfig{file=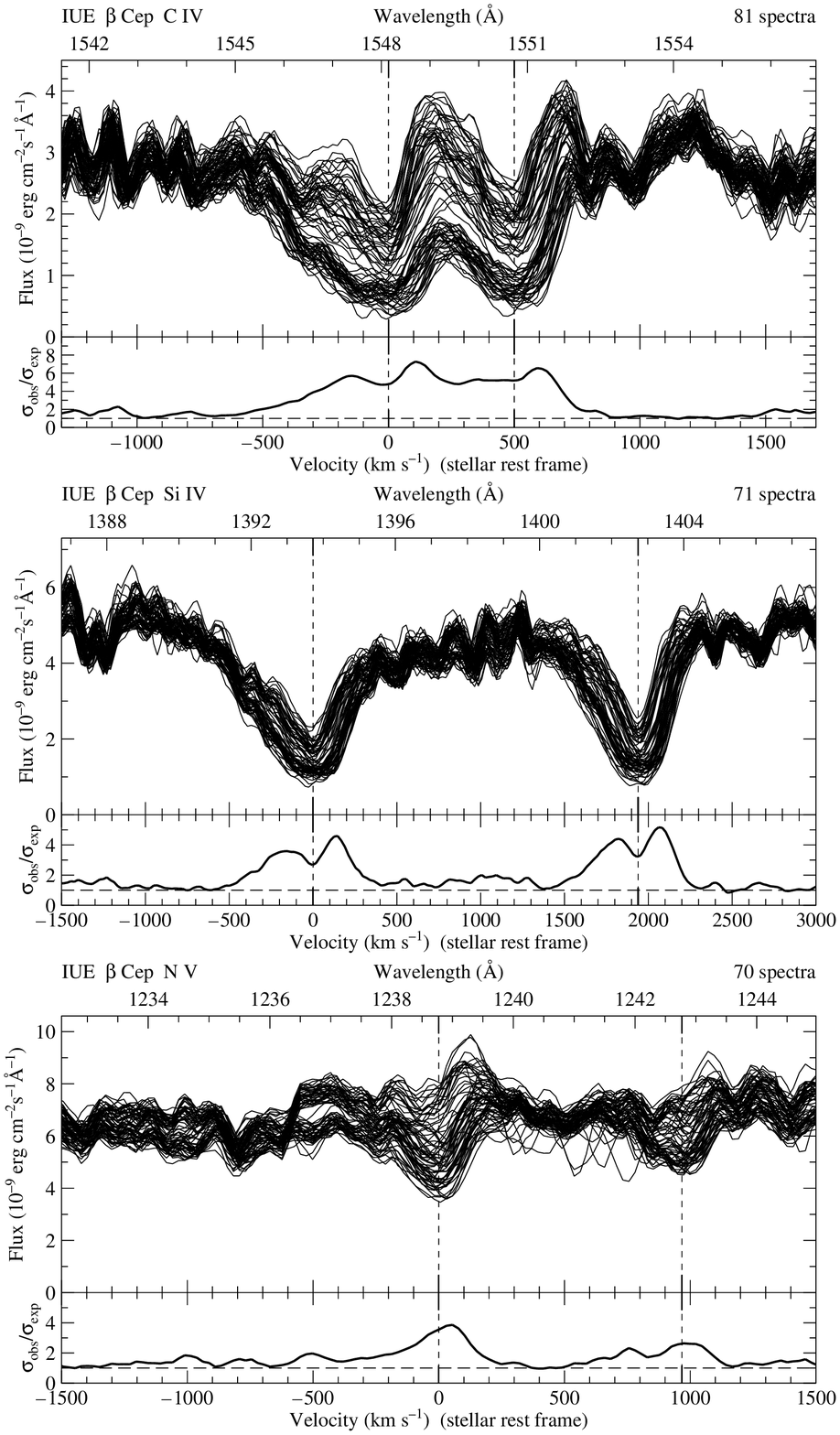,width=\columnwidth,clip=}
\caption{Representative wind profiles from IUE spectra showing the
typical variation over a 12-day cycle, which is very similar to the type of variation
observed in other magnetic B stars but unlike the variations observed in O
stars. Top: \ion{C}{iv}. Middle: \ion{Si}{iv}. Bottom: \ion{N}{v}.  The two doublet rest wavelengths are indicated by vertical dashed lines. Top scales: Wavelength. Bottom scales: Velocity relative to the stellar rest frame.
In each panel the lower part displays the significance of the variability, as the ratio of the measured to the expected variances.}
\label{fig:uv}
\end{center}
\end{figure}

From IUE spectra, it is known that the UV stellar wind lines of \ion{C}{iv},
\ion{Si}{iv}, and \ion{N}{v}  show a very clear
12-day periodicity. In Fig.~\ref{fig:uv}, we show the behaviour of
these doublet UV profiles, along with the significance of variability.
We used here the noise model for high-resolution IUE spectra from 
\cite{henrichs:1994}, with parameters $A$ = 18, which represents the
average signal-to-noise ratio, and $B = 2 \times 10^{-9}$ erg cm$^{-2}$s$^{-1}$\AA$^{-1}$, which represents the
average flux level.  Note that the outflow velocity exceeds $-$600
km~s$^{-1}$. The real terminal velocity of the wind cannot be measured with this relatively low S/N ratio.
 We also note that this type of variability is very similar
to profile variations of other magnetic B stars (see Sect.\ 1), which is unlike what
is observed in O stars \citep[e.g.][]{kaper:1996}. The discrete absorption 
components in O stars march from low to high (negative) velocity in the absorption part of the P Cygni profile on a rotational timescale. The origin of this behaviour in O stars is not well understood, although some promising contenders have been identified (e.g., Corotating Interacting Regions).

We measured the 
equivalent width (EW) of the  \ion{C}{iv} 1548, 1551 lines in the velocity range of [$-$700, 800] km s$^{-1}$,
after normalising 81 out of 88 available IUE spectra between 1978 and 1995
to the same continuum around the \ion{C}{iv} line and dividing each spectrum
by the average of the normalised spectra. (The remaining 7 spectra were not well calibrated.)
The error bars are calculated
following \cite{chalabaev:1983}. The resulting EW values are
plotted as a function of phase in Fig.~\ref{fig:ewuvb} (2nd panel).
The same procedure was carried out for the lines of \ion{Si}{iv} (3rd panel)
and \ion{N}{v} (lower panel),
where we used 70 spectra with the highest quality and integrated over the intervals
[$-$600, 2500] and [$-$600, 1200] km s$^{-1}$, respectively.

We used a superposition of two sine waves to fit the \ion{C}{iv} EW data.
The result is obtained with a least-squares method, which uses weights equal to $1/\sigma^2$
(with 1$\sigma$ the individual error bars) assigned to each datapoint.
With user-supplied initial starting values for the free parameters, a steepest descent technique then searched for the lowest minimum of the $\chi^2$.  The variance matrix provides the formal
errors in the parameters, as seen in the following function:
$f(t)= a+b(\sin(2 \pi(t/P+d)))+e(\sin(2\pi(t/(P/2)+f)))$.

The results of the best solution with a reduced $\chi^2 =0.52$ are: $a =
2.41 \pm 0.03$, $b=0.60\pm0.05$, $d= 0.308\pm 0.009$, $e= 1.77 \pm0.04$,
$f = 0.84\pm0.01$, and a period $P = 12.00075 \pm 0.00011$ d. The very high
precision of less than 10 seconds in the period is thanks to extended coverage over almost 500
cycles. All doublet profiles of \ion{C}{iv}, \ion{Si}{iv}, and \ion{N}{v} 
are modulated with this same period, which is identified with the rotation
period of the star.  With this analytic description, the epoch of minimum 
EW could be derived mathematically. We derived the ephemeris for the deepest
minimum (i.e.\ maximum emission), which we define as the zero phase of the
rotation. We find

\begin{eqnarray}
\nonumber T({\rm EW_{min}}) = &{\rm HJD}\ 2449762.050\pm0.063 &\\
&+\ n \times(12.00075 \pm 0.00011)&
\label{eq:uvphase}
\end{eqnarray}

with $n$ defined as the number of cycles. The reference date HJD 2449762.050 is at the EW minimum that is closest to the middle of the IUE observations.

The phase difference between the maxima in the fitted EW curve is $\Delta\phi_{\rm max} = 0.527 \pm 0.003$, whereas the difference between the minima is $\Delta\phi_{\rm min} = 0.504 \pm 0.002$.  The difference between these two values, and the unequal depths clearly indicates that the wind must be asymmetric with respect to the magnetic equator. For a banded oblique rotator \citep{shore:1987a}, the value of $\Delta\phi$ is directly related to the asymmetry in the magnetic modulation, which will be compared in Sect. \ref{subsec:comparison}.

\subsection{Magnetic properties}

A CLEAN analysis of all magnetic data reveals only one significant period of $P$ = 11.9971 $\pm$ 0.0078 d. This period is within the uncertainties equal to the period derived from the much more accurate UV data. We therefore use the UV period in our further analysis. 
We fitted a cosine function of the following form to the 130 $B_\ell$
datapoints with  $1/\sigma^2$ error bars as weights:

\begin{equation}
B_\ell (t) = B_0+ B_{\rm max} \cos (2\pi(\frac{t}{12.00075}+\phi))
\end{equation}

\begin{figure}[t!]
\begin{center}
\epsfig{file=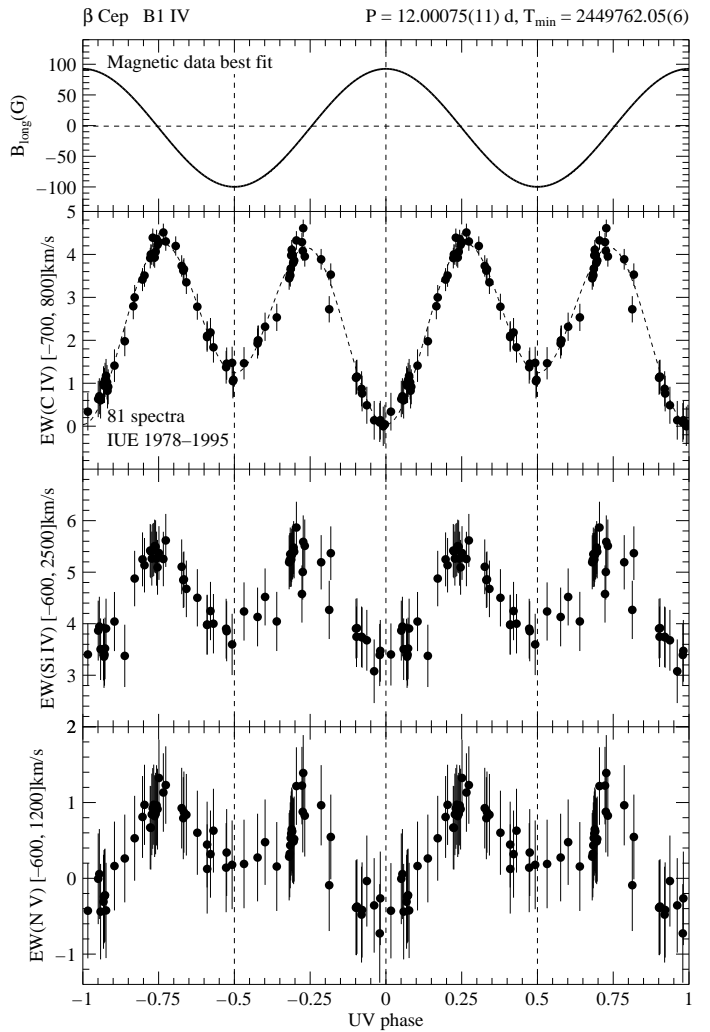,width=\columnwidth,clip=}
\caption{Phase behaviour of the stellar wind lines and magnetic variation. Upper panel: best sinusoid fit to the magnetic data 
of $\beta$~Cep as a function of the UV phase, using a fixed period of 12.00075 d, as derived from UV data. Two rotational periods are displayed. Vertical dashed lines at phases 0.5 and 1 serve as a reference for comparison with the stellar wind behaviour.
Lower panels, respectively, from top to bottom:
equivalent width of the \ion{C}{iv}, \ion{Si}{iv}, and \ion{N}{v}  stellar wind lines
measured in IUE spectra taken during 16 years as a function of phase, which is
calculated with Eq.~\ref{eq:uvphase}. The deepest minimum, defined as phase 0,
corresponds to the maximum emission, which occurs when the magnetic North pole is pointing most to the observer.  Note that there is no significant difference in
zero phase between the UV and magnetic data but that the field crosses zero
slightly besides the EW maxima, as discussed in Sect. \ref{subsec:comparison}}
\label{fig:ewuvb}
\end{center}
\end{figure}

in which $t$ was taken relative to the first observation.
The best-fit values are $B_0$ = $-$6 $\pm$ 3 G, $B_{\rm max}$ = 97 $\pm$ 4 G,
and $\phi = 0.4759 \pm 0.0079$ with a reduced $\chi^2$ = 2.3.

With the derived phase, we find the maximum value of the field strength for the
ephemeris:
\begin{equation}
T({\rm B_{\rm max}}) = {\rm HJD }\ 2452366.25 \pm 0.10 + n\times12.00075.
\label{eq:bphase}
\end{equation}
The reference date HJD 2452366.25 is given at the maximum field closest to the middle of the magnetic measurements, which extend over 200 cycles.

When omitting the 21 data points of the 2000 dataset because of the potential fringing problem mentioned above, we obtain $B_0$ = $-$11 $\pm$ 3 G, $B_{\rm max}$ = 101 $\pm$ 4 G, $\phi = 0.4913 \pm 0.0077$ with a reduced $\chi^2$ = 2.1, and a reference date of HJD 2452366.18 $\pm$ 0.09, which is hardly different from the fit with all data. Given the very small differences, we consider the fit above with all 130 datapoints still as the best.

\begin{figure}[t!]
\begin{center}
\epsfig{file=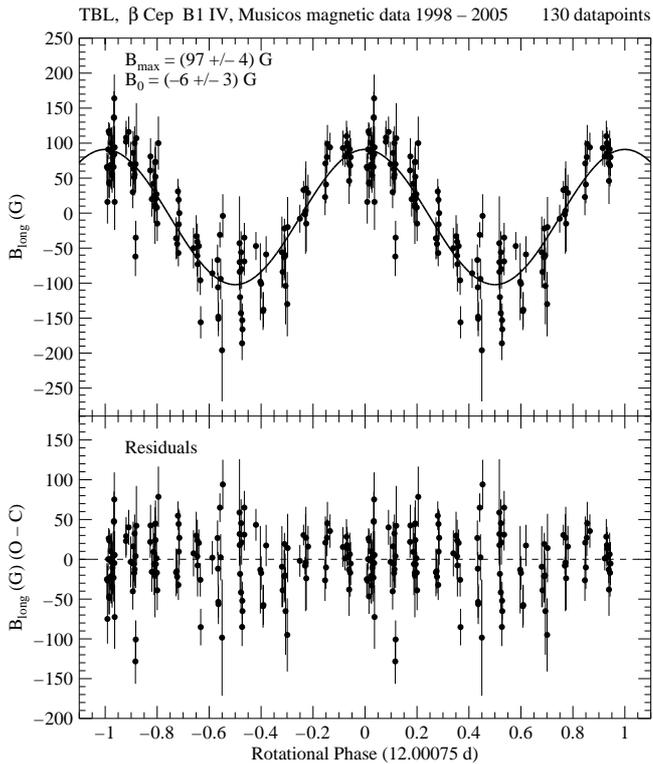,width=\columnwidth,clip=}
\caption{{\sl Top:} Overplot of all magnetic data folded with the rotational
period of 12.00075 d. Two rotational periods are displayed. The drawn curve is the best-fit cosine with amplitude 97 G and
offset $-$6 G. {\sl Bottom:} Overplot of residuals (O $-$ C) of all magnetic data folded with the rotational period.
}
\label{fig:bphaser}
\end{center}
\end{figure}

In Fig.~\ref{fig:bfield}, we have drawn a sine wave with this period and
phase through the early magnetic measurements of December 1998 until July 1999.
Figs.~\ref{fig:ewuvb} and \ref{fig:bphaser} show all 130 datapoints folded with the rotational period
and an overplot of the best-fit cosine curve. In Fig.~\ref{fig:bphaser}, the residuals from the fit are displayed.
No obvious discrepancies are emerging with the limited accuracy of the present data.

\subsection{Comparison between magnetic and wind properties}
\label{subsec:comparison}

A comparison to the phase of the UV data (Eq.~\ref{eq:uvphase}) shows that
a deep EW minimum is predicted at HJD 2452366.21 $\pm$ 0.04, which is,
within the uncertainties, identical to the phase of maximum (positive)
magnetic field. In Fig.~\ref{fig:ewuvb} (2nd panel), we have drawn a sine
wave with the best fit parameters through the
values of the magnetic field strength and using the phase from the UV period. It is
clear from the figure that the phase of minima of the stellar wind
absorption (i.e.\ maximum emission)  coincides very well with the extremes of
the magnetic field, and that the maximum wind absorption coincides with a
field strength zero. This is compatible with an oblique rotator model in which the wind 
outflow is much enhanced in the plane of the magnetic equator.
It is of interest to note that the result of $B_0$ = $-$6 $\pm$ 3 G implies that the asymmetry
relative to zero must be small, although the shallower second EW
minimum implies an asymmetric wind. This can be used to put constraints on the geometry of the field. As outlined by \cite{shore:1987a}, the ratio of the extremes of the magnetic modulation $r=B_{\rm max}/B_{\rm min}$ is related to the phase difference $\Delta\phi$ between the maxima or minima of the EW modulation by $r = (u-1)/(u+1)$ with $u=cos(\Delta\phi/2)$. The values for $\Delta\phi$ in the previous section imply that $B_0 = 8.0 \pm 1.2$ G when de maxima are used, and that $B_0 = 0.6 \pm 0.7 $ G when the minima are considered. A geometry that is compatible with the phase difference between the minima in the EW (with maximum positive field value) is therefore slightly favored.  It is clear, however, that more magnetic data are needed to confirm any asymmetry.

For a dipolar field, the ratio  $r =
B_{\rm min}/B_{\rm max}$ is also related to the inclination angle, $i$, and the
angle of the magnetic axis relative to the rotation axis, $\beta$, according to $r$ =
cos($\beta +i$)/cos($\beta -i$) \citep{preston:1967}. 
We obtain $r=-1.12\pm0.01$, which can put
further constraints upon the angles $\beta$ and $i$. For $i$ near $60^{\circ}$, it follows that $\beta$ close to $96^{\circ}$. 
With a projected rotational velocity of 27(3) km~s$^{-1}$ (as
derived by \cite{telting:1997} from a pulsation mode analysis), the
rotation period requires a radius of 7.4 $\pm$ 0.8 R$_{\odot}$, which is in excellent agreement with the value 7.5 $\pm$ 0.7 R$_{\odot}$ that is based on \cite{remie:1982}. These authors derived the angular size from the integrated total flux and infrared flux. A further discussion of the stellar parameters
and the best-fit results of the angles $i$ and $\beta$ based on line fits to $I$ and
$V$ is given by \cite{donati:2001}.

\subsection{Pulsation period and system velocity}

The measured radial velocities of the star are given in column 9 in
Table~\ref{tab:results}. For the calculation of the phase in
heliocentric radial velocity
due to the radial mode of the pulsation, we used the ephemeris for the
expected maximum from \cite{pigulski:1992} with
$P$ = 0.1904852~d and $T_{\rm max}$ = 2413499.5407 (column 7 in
Table~\ref{tab:results}).

\begin{figure}[b!]
\begin{center}
\epsfig{file=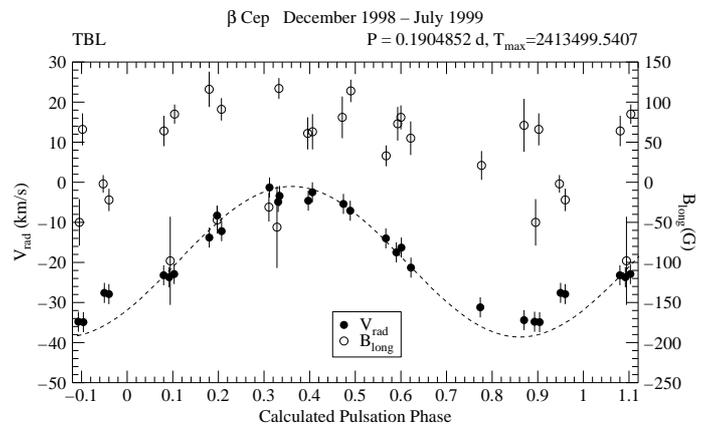,width=\columnwidth,clip=}
\caption{Derived radial velocity (filled symbols, scale on the left) and
magnetic field strength (open symbols, scale on the right) as a function of
pulsation phase. As expected, no correlation between the two quantities is
present in the data: at several occasions very different magnetic values are
measured at a given pulsation phase.  The observed system velocity and the difference between the observed and calculated phase of the maximum
radial velocity confirm the predicted values for the star in its binary
orbit near periastron passage.}
\label{fig:pulseb} 
\end{center}
\end{figure}

In Fig.~\ref{fig:pulseb}, we plotted the derived radial velocity 
with the magnetic field strength as a function of the calculated pulsation
phase for the first 23 measurements, which covered 7 months in 1998 and 1999.
Taking all the data together would not make sense because of the expected systemic
velocity of the star in its 90-year orbit, especially since the star was very
near its periastron passage (see below).
From the figure, it is clear that there is no correlation between the
pulsation phase and the longitudinal component of the magnetic field, as expected. The correlation coefficient for a sine curve is only 0.11.

\begin{figure}[!hb]
\begin{center}
\epsfig{file=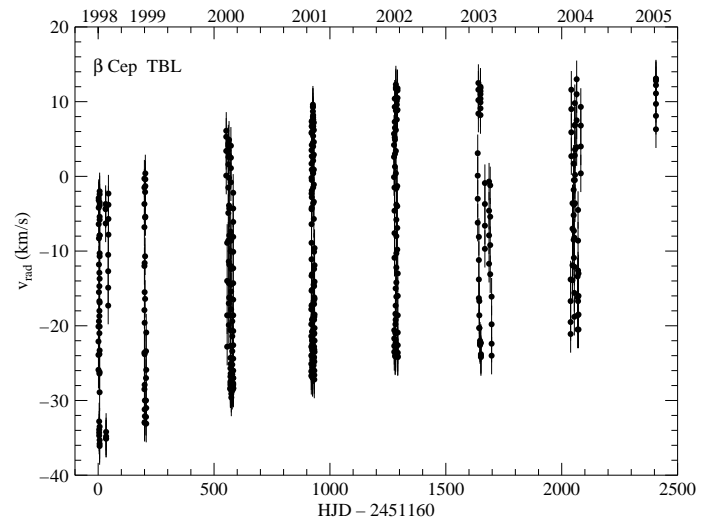,width=\columnwidth,clip=}
\caption{Radial velocities as measured from all spectra from 1998-2005.}
\label{fig:vradall}
\end{center}
\end{figure}

\begin{table*}
\begin{center}
\caption[]{Results from cosine fits using Eq.\  \ref{eq:cosinepuls} of radial velocity data of all 477 individual subexposures, subdivided in yearly averages. }
\begin{tabular}{lcccccc}
\hline
\hline
Data set     & Av.\ HJD & Coverage & Nr     &  \multicolumn{1}{c}{$\gamma$}      & Phase $\phi$ & O $-$ C\\
           & $-$2450000 & (days)   & points & \multicolumn{1}{c}{(km s$^{-1}$)}  &       & (days)\\
\hline
1998-99 Jan  & 1182.81 &  \042.9 & \060 &  $-$19.15$\pm$0.09 &    0.6781 $\pm$0.0012   &$-$0.1292\\
1999 Jun     & 1363.46 & \0\07.9 & \034 &  $-$17.07$\pm$0.19 &    0.7215 $\pm$0.0018   &$-$0.1374\\
2000 Jun     & 1727.64 &  \030.0 & \088 &  $-$12.43$\pm$0.12 &    0.7480 $\pm$0.0015   &$-$0.1429\\
2001 Jun     & 2086.59 &  \014.0 &  123 & \0$-$9.25$\pm$0.07 &   0.75844 $\pm$0.00095  &$-$0.1445\\
2002 Jun     & 2444.55 &  \016.1 & \072 & \0$-$6.85$\pm$0.08 &    0.7601 $\pm$0.0012   &$-$0.1448\\
2003 Jun-Aug & 2827.58 &  \059.9 & \048 & \0$-$6.42$\pm$0.20 &    0.7521 $\pm$0.0025   &$-$0.1433\\
2004 Jun-Nov & 3267.90 &   138.8 & \052 & \0$-$5.05$\pm$0.14 &    0.7499 $\pm$0.0013   &$-$0.1428\\
\hline
\hline
\end{tabular}
\label{tab:vradpuls}
\end{center}
\end{table*}

Each magnetic measurement consists of four subexposures for each of which
we determined the radial velocity. (Only the average value per 4 subexposures is given in 
Table~\ref{tab:results}.)
We also included spectra from incomplete sets, which made a total of 477 data points, as shown in Fig. \ref{fig:vradall}.
We divided this dataset into logical subsets, with a coverage of about 0.5 to 3 weeks each year (see column 3 in Table \ref{tab:vradpuls}) and performed cosine fits for each subset with the function:

\begin{equation}
\label{eq:cosinepuls}
v_{\rm rad}(t) = \gamma +A\cos(2\pi((t-t_0)/P+\phi)).
\end{equation}

The ephemeris and the period $P$ from \cite{pigulski:1992} were used to derive the phase 
$\phi$ and the delay of the maximum of the radial velocity curve of the pulsation (O $-$ C), which is attributed to the light time effect.
 The results are given in Table \ref{tab:vradpuls}. A best fit through the values of the system velocity $\gamma$ of Table \ref{tab:vradpuls} yielded the orbital parameters listed in Table \ref{tab:orbit}. We kept $T_{0}$, the passage of periastron, constant in this fit. Our orbital parameters agree fairly well with those derived by \cite{pigulski:1992}. With only 7 data points and 5 free parameters the error bars are expectedly large for some of the parameters. We note that $\omega$ becomes 198$^\circ$ $\pm$ 2$^\circ$ if all other values are kept constant, and that the error bar is derived by letting $\chi^2$ increase by unity. 
Our values of the delays are also in good agreement with the expected phase delay , which is caused by the light-time effect in the binary orbit \citep[see][ their Fig.\ 1]{pigulski:1992}.

\begin{table}
\begin{center}
\caption[]{New orbital parameters with 1$\sigma$ errors for the $\beta$ Cep system, which is based on radial velocity measurements in this paper.
For comparison, the values are listed as given by \cite{pigulski:1992} (PB92), which are based on radial velocities, and \cite{andrade:2006} (A06), which are based on interferometric data.}
\begin{tabular}{llll}
\hline
\hline
Parameter              & This paper       & PB92   & A06\\
\hline
$P_{\rm orb}$ (yr)     & 84.5 $\pm$ 0.6   & 91.6 $\pm$ 3.7   & 83 $\pm$ 9\\
$e$                    & 0.74 $\pm$ 0.11  & 0.65 $\pm$ 0.03  & 0.732 $\pm$ 0.16\\
$\omega$($^\circ$)     & 217 $\pm$ 61     & 194 $\pm$ 4      & 194.6 $\pm$ 2.5\\
$T_0$                  & 1914.6 (fixed)   & 1914.6 $\pm$ 0.4 & 1997.99 $\pm$ 0.10\\
$K_1$ (km s$^{-1}$)    & 8.2 $\pm$ 1.3    & 8.0 $\pm$ 0.5    & \\
$\gamma$ (km s$^{-1}$) & $-$7.5 $\pm$ 7.5 & $-$6.6 $\pm$ 0.4 & \\
$i$($^\circ$)          &                  &                  & 87.3 $\pm$ 1.5\\ 
$a$($''$)              &                  &                  & 0.195 $\pm$ 0.08\\  
$\Omega$($^\circ$)     &                  &                  & 46.4 $\pm$ 1.5\\
\hline
\hline
\end{tabular}
\label{tab:orbit}
\end{center}
\end{table}

\begin{figure}[!ht]
\begin{center}
\epsfig{file=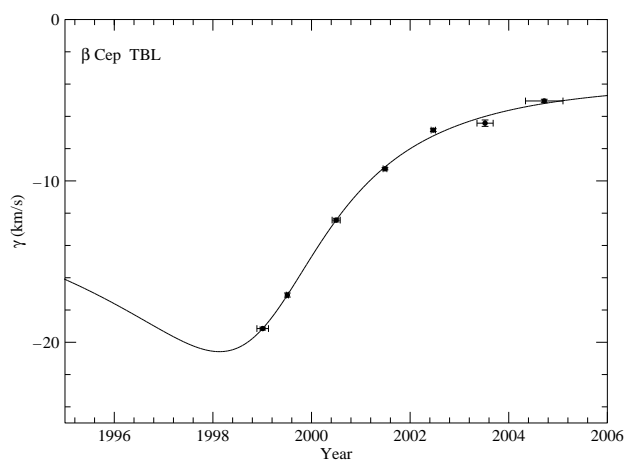,width=0.9\columnwidth,clip=}
\caption{Observed system velocity $\gamma$ overplotted with the orbital solution from Table ~\ref{tab:orbit}.
}
\label{fig:vfit}
\end{center}
\end{figure}

In the binary system, the companion has been resolved by interferometry.
New speckle interferometric measurements by
\cite{balega:2002} yielded a separation of 38$\pm$2 mas at the epoch 1998.770 (i.e.\ just preceding our first observation) at a position angle of 228.6$^{\circ}$, whereas
the position angle was 49.3$^{\circ}$ \citep{hartkopf:1992} with separation 50 mas 8 years earlier. This
shows that the companion had passed a minimum radial velocity before
our first observation. The binary period is therefore
likely a little shorter than 85 years, based on the previous periastron passage
in 1914.6 $\pm$ 0.4.  This agrees with \cite{hadrava:1996},
who concluded that the periastron passage should be closer to 1996 than to 2006, as predicted
by \cite{pigulski:1992}.  Our orbital solution points to the same conclusion. A more accurate solution  
was achieved by \cite{andrade:2006} with a new analysis, based on all available interferometric measurements between 1971.48 and 1998.77, which included the periastron passage. 
These results, given in column 3 Table \ref{tab:vradpuls}, are in very good agreement of the findings of this paper.

\section{Conclusions and discussion}

We have unambiguously found a varying weak magnetic field
in $\beta$~Cep, which is consistent with an oblique dipolar magnetic rotator
model with a rotation period of 12 days. Contrary to what is found in models by \cite{brown:1985}, \cite{shore:1987a}, 
and \cite{shore:1987b}, the UV wind-line absorption
in $\beta$~Cep is at a maximum if the Earth is in the magnetic equatorial
plane. Model MHD calculations of the outflow and the resulting UV wind
lines with a SEI-based code that were applied to $\beta$~Cep are presented by \cite{Schnerr:2007}. 
They found that the observed variability of the wind lines could be qualitatively reproduced for simple phenomenological models with enhanced density in the magnetic equator (presumably due to
magnetic channeling of the wind).
However, significant
differences are encountered when full 2D-MHD models were used to determine the geometry of the stellar wind. The authors ascribe this effect to X-ray ionisation that has not been included.
 
We emphasize that we have only measured the longitudinal component of the
magnetic field, (i.e., the component in the line of sight, which is taken averaged
over the stellar disk).  The intensity at the magnetic poles must be
stronger.  For a perpendicular magnetic rotator the polar field of a dipole
is 3.2$\times B_{\ell, {\rm max}}$ \citep{schwarzschild:1950}, or about
300 G. However, the EW curve in the stellar wind lines has
two unequal maxima at epochs when the projected field is strongest, which suggests
that there should be a slight asymmetry present.  This could of course be
due to a slightly different geometry (off-centered dipole or higher-order
fields) at the two hemispheres, which can easily be hidden in the observed
field strength that is the integrated value over the visible surface. 
The EW maxima of the UV wind lines occur slightly off the phases 0.25 and 0.75, when the magnetic values cross zero, 
which also indicates some asymmetry.

We note that ths configuration found here favors magnetic braking as discussed
by \cite{donati:2001}, although the origin of the present slow rotation rate remained unclear.
 They calculated the characteristic spindown time to be 110 My for a mass-loss rate of 2.7 
$\times 10^{-10}$ M$_{\odot}$y$^{-1}$ at an Alfv\'{e}n radius of 9 R$_{\ast}$, which is 
about an order of magnitude longer than the age of the star. Applying the work by \cite{ud-Doula:2008} to $\beta$~Cep gives a similar result. However, this may be an upper limit, as their model is applied to a pure dipole field aligned with the rotation axis, unlike the configuration in $\beta$~Cep, which is an oblique rotator.
We also must consider the possibility, as mentioned above, that the field is not dipole-like but resembles 
a split monopole, in which case the estimated spindown time can be as much as two orders of magnitudes shorter, 
which is of the order of 1 My (i.e., clearly within the lifetime of the star \citep{ud-doula:2009}). 

It is also interesting to note that the mode splitting due to the rotation
is clearly present in the pulsation properties \citep{telting:1997}.
It would be worth examining whether the presence of the magnetic field can also
be traced back in the pulsation modes. If so, this will give a strong
constraint on the evolutionary status of $\beta$~Cep. In this regard, it is clear that the analysis as presented by \cite{shibahashi:2000} should be revised, as it was based on a rotational period of 6 days rather than 12 days.

The X-ray flux of $\beta$~Cep was detected by the {\it Einstein} Observatory \citep{grillo:1992} and later by the ROSAT satellite \cite{berghofer:1996}. After the discovery of the magnetic field, the (static) magnetically confined shock (MWCS) model by \cite{babel:1997} was used by \cite{donati:2001} to make specific predictions of the X-ray behaviour as a function of rotational phase.  Dynamical modelling of the interaction of the wind and the magnetic field by \cite{ud-doula:2002}, \cite{gagne:2005} and \cite{townsend:2005} showed, however, that an optically thick cool disk, as predicted by the MWCS model, would not form around $\beta$ Cep.  The lack of the predicted rotational modulation in the flux measured with the {\it Chandra} and XMM/{\it Newton} observatories analysed by \cite{favata:2009} confirmed the absence of such a static disk. The analysis of line ratios in the X-ray spectra allowed these authors to put constraints on the distance from the star where the relatively low-temperature X-ray plasma is confined, which is typically a few stellar radii, with a stratified temperature distribution.  They also discuss the significant differences in X-ray and wind-confinement properties between $\beta$ Cep and the more massive magnetic O4 star 
$\theta^1$~Ori C, for which the MWCS model is compatible with the observed X-ray characteristics.

The close environment of $\beta$~Cep was recently investigated with interferometry by \cite{nardetto:2011} with a spatial resolution of 1 mas. 
They concluded that a circular ring (instead of a uniform disk) gives the best fit with a geometry and orientation similar to the magnetic equator, as described above, which is a remarkable result. They determined this result
from model fits of the resolved large-scale structure around the star at two different rotational phases (0.47 and 0.17 with our ephemeris, which is about 0.03 lower than used in their paper).  Their favored geometry gives an inner ring diameter of about $74\ R_{*}$, a width of $5\ R_{*}$, and a position angle of $60^{\circ}$. As discussed by these authors, this ring should be optically thin in the X-ray band and is also consistent with the absence of X-ray modulation reported by \cite{favata:2009}. The radius of this ring is, however, larger than the region constrained by \cite{favata:2009}.

Several other issues are still to be solved. First of all, why does  $\beta$~Cep have
a magnetic field? This star does not belong to the
helium-peculiar stars \citep{rachkovskaya:1990}, which are known to have
strong magnetic fields (see for instance \cite{bohlender:1987}). We did not find any 
variability in the wind-sensitive \ion{He}{i} 1640 line,  although not much can be expected to show up in this very weak line with the very poor S/N ($< 10$ in this wavelength region).     \cite{gies:1992}
 note that $\beta$~Cep is N enriched, a property that this star shares with
other magnetic B stars, as confirmed by \cite{morel:2006} and \cite{nieva:2012}. Enrichment of nitrogen in
the atmosphere of a B star is apparently a strong indirect indicator of a surface magnetic field \citep{henrichs:2005}. 
The presence of a magnetic field could cause these anomalies by inhibiting mixing in the interior, but no specific models have been developed to explain the N enrichment.  Theoretical
predictions for the nitrogen enrichment at the surface of B stars that host a large-scale,
dipolar field have been presented by \cite{meynet:2011}. \cite{nieva:2012} considered the overabundance of nitrogen as a result of mixing of CN-recycled material into the stellar atmosphere. \cite{wade:2007} proposed a simple mechanism for why some A and B stars have magnetic fields.

Another point of concern is that we have fitted a simple sine curve through the magnetic data.  This
is obviously a first approximation, and when more accurate measurements become
available, a search for deviations from a sine curve, as is found for most
magnetic stars, can be done.

Lastly, one may wonder whether other pulsating B stars may show the same type of wind variability, which is
specific for magnetic stars. An elaborated search by Henrichs and ten Kulve (in 
preparation) in the complete sample of IUE spectra of 395 B stars provided the best 
possible statistics to date. None of these stars (besides the well known 
magnetic Bp stars) except for \object{$\beta$ Cep}, \object{V2052 Oph} \citep{neiner:2003b, briquet:2012}, \object{$\zeta$ Cas} \citep{neiner:2003a}, 
and the possibility of \object{$\sigma$ Lup} \citep{henrichs:2012} showed a similar periodic pattern. This must be regarded as a lower limit, since 
only very limited timeseries with poor coverage exist in most cases, 
which makes it impossible to quantify the real occurrence. \cite{hubrig:2011} searched in a number of pulsating B stars for magnetic fields. \cite{silvester:2009} concluded from a study of 30 stars that magnetism is not common among pulsating B stars. Recently, \cite{petit:2013}  discussed the magnetic environments of all discovered massive magnetic stars, among which $\beta$~Cep does magnetically not stand out.
The star $\beta$~Cep hence appears to be one of the
very few stars in its class that shows the type of strong wind variability described here. In this respect $\beta$~Cep is an exceptional $\beta$~Cephei star. 

\begin{acknowledgements}
We thank D.\ Lennon, G.\ Mathys, S.\ Solanki, J.\ Telting, and A.\ ud-Doula
for discussions and constructive comments. We also thanks an anonymous referee for useful remarks and comments. HFH thanks his coauthors for their patience and S. Hubrig for encouragement to finish this work.
The helpful assistance and support of the observatory staff members
at TBL, GSFC, and Vilspa (in particular, the late Dr. Willem Wamsteker) 
is well remembered and fondly acknowledged.
JDJ acknowledges support from the Netherlands Foundation for Research in
Astronomy (NFRA) with financial aid from the Netherlands Organization for
Scientific Research (NWO) under project 781-71-053.  GAW acknowledges
support from the Natural Sciences and Engineering Council of Canada (NSERC).
\end{acknowledgements}

\bibliographystyle{aa}
\bibliography{../../../references}

\begin{thebibliography}{72}
\expandafter\ifx\csname natexlab\endcsname\relax\def\natexlab#1{#1}\fi

\bibitem[{{Abt} {et~al.}(2002){Abt}, {Levato}, \& {Grosso}}]{abt:2002}
{Abt}, H.~A., {Levato}, H., \& {Grosso}, M. 2002, \apj, 573, 359

\bibitem[{{Andrade}(2006)}]{andrade:2006}
{Andrade}, M. 2006, IAU Inf. Circ.\ Comm.\ 26, 158, 158, 3

\bibitem[{{Babel} \& {Montmerle}(1997)}]{babel:1997}
{Babel}, J. \& {Montmerle}, T. 1997, \aap, 323, 121

\bibitem[{{Balega} {et~al.}(2002){Balega}, {Balega}, {Hofmann}, {Maksimov},
  {Pluzhnik}, {Schertl}, {Shkhagosheva}, \& {Weigelt}}]{balega:2002}
{Balega}, I.~I., {Balega}, Y.~Y., {Hofmann}, K.-H., {et~al.} 2002, \aap, 385,
  87

\bibitem[{{Barker} {et~al.}(1982){Barker}, {Brown}, {Bolton}, \&
  {Landstreet}}]{barker:1982}
{Barker}, P.~K., {Brown}, D.~N., {Bolton}, C.~T., \& {Landstreet}, J.~D. 1982,
  in Advances in Ultraviolet Astronomy, ed. Y.~{Kondo}, 589--592

\bibitem[{{Baudrand} \& {Bohm}(1992)}]{baudrand:1992}
{Baudrand}, J. \& {Bohm}, T. 1992, \aap, 259, 711

\bibitem[{{Bergh\"{o}fer} {et~al.}(1996){Bergh\"{o}fer}, {Schmitt}, \&
  {Cassinelli}}]{berghofer:1996}
{Bergh\"{o}fer}, T.~W., {Schmitt}, J.~H.~M.~M., \& {Cassinelli}, J.~P. 1996,
  \aaps, 118, 481

\bibitem[{{Bohlender} {et~al.}(1987){Bohlender}, {Landstreet}, {Brown}, \&
  {Thompson}}]{bohlender:1987}
{Bohlender}, D.~A., {Landstreet}, J.~D., {Brown}, D.~N., \& {Thompson}, I.~B.
  1987, \apj, 323, 325

\bibitem[{{Briquet} {et~al.}(2012){Briquet}, {Neiner}, {Aerts}, {Morel},
  {Mathis}, {Reese}, {Lehmann}, {Costero}, {Echevarria}, {Handler}, {Kambe},
  {Hirata}, {Masuda}, {Wright}, {Yang}, {Pintado}, {Mkrtichian}, {Lee}, {Han},
  {Bruch}, {De Cat}, {Uytterhoeven}, {Lefever}, {Vanautgaerden}, {de Batz},
  {Fr{\'e}mat}, {Henrichs}, {Geers}, {Martayan}, {Hubert}, {Thizy}, \&
  {Tijani}}]{briquet:2012}
{Briquet}, M., {Neiner}, C., {Aerts}, C., {et~al.} 2012, \mnras, 427, 483

\bibitem[{{Brown} {et~al.}(1985){Brown}, {Shore}, \& {Sonneborn}}]{brown:1985}
{Brown}, D.~N., {Shore}, S.~N., \& {Sonneborn}, G. 1985, \aj, 90, 1354

\bibitem[{{Catala} {et~al.}(1993){Catala}, {Foing}, {Baudrand}, {Cao}, {Char},
  {Chatzichristou}, {Cuby}, {Czarny}, {Dreux}, {Felenbok}, {Floquet}, {Geurin},
  {Huang}, {Hubert-Delplace}, {Hubert}, {Huovelin}, {Jankov}, {Jiang}, {Li},
  {Neff}, {Petrov}, {Savanov}, {Shcherbakov}, {Simon}, {Tuominen}, \&
  {Zhai}}]{catala:1993}
{Catala}, C., {Foing}, B.~H., {Baudrand}, J., {et~al.} 1993, \aap, 275, 245

\bibitem[{{Chalabaev} \& {Maillard}(1983)}]{chalabaev:1983}
{Chalabaev}, A. \& {Maillard}, J.~P. 1983, \aap, 127, 279

\bibitem[{{Donati} {et~al.}(1999){Donati}, {Catala}, {Wade}, {Gallou},
  {Delaigue}, \& {Rabou}}]{donati:1999}
{Donati}, J.-F., {Catala}, C., {Wade}, G.~A., {et~al.} 1999, \aaps, 134, 149

\bibitem[{{Donati} {et~al.}(1997){Donati}, {Semel}, {Carter}, {Rees}, \&
  {Collier Cameron}}]{donati:1997}
{Donati}, J.-F., {Semel}, M., {Carter}, B.~D., {Rees}, D.~E., \& {Collier
  Cameron}, A. 1997, \mnras, 291, 658

\bibitem[{{Donati} {et~al.}(2001){Donati}, {Wade}, {Babel}, {Henrichs}, {de
  Jong}, \& {Harries}}]{donati:2001}
{Donati}, J.-F., {Wade}, G.~A., {Babel}, J., {et~al.} 2001, \mnras, 326, 1265

\bibitem[{{Favata} {et~al.}(2009){Favata}, {Neiner}, {Testa}, {Hussain}, \&
  {Sanz-Forcada}}]{favata:2009}
{Favata}, F., {Neiner}, C., {Testa}, P., {Hussain}, G., \& {Sanz-Forcada}, J.
  2009, \aap, 495, 217

\bibitem[{{Fischel} \& {Sparks}(1972)}]{fischel:1972}
{Fischel}, D. \& {Sparks}, W.~M. 1972, in The scientifique results from the
  Orbiting Astronomical Observatory (OAO-2), NASA SP-310, 475

\bibitem[{{Frost} \& {Adams}(1903)}]{frost:1903}
{Frost}, E.~B. \& {Adams}, W.~S. 1903, \apj, 17, 150

\bibitem[{{Gagn{\'e}} {et~al.}(2005){Gagn{\'e}}, {Oksala}, {Cohen}, {Tonnesen},
  {ud-Doula}, {Owocki}, {Townsend}, \& {MacFarlane}}]{gagne:2005}
{Gagn{\'e}}, M., {Oksala}, M.~E., {Cohen}, D.~H., {et~al.} 2005, \apj, 628, 986

\bibitem[{{Gezari} {et~al.}(1972){Gezari}, {Labeyrie}, \&
  {Stachnik}}]{gezari:1972}
{Gezari}, D.~Y., {Labeyrie}, A., \& {Stachnik}, R.~V. 1972, \apjl, 173, L1

\bibitem[{{Gies} \& {Lambert}(1992)}]{gies:1992}
{Gies}, D.~R. \& {Lambert}, D.~L. 1992, \apj, 387, 673

\bibitem[{{Grillo} {et~al.}(1992){Grillo}, {Sciortino}, {Micela}, {Vaiana}, \&
  {Harnden}}]{grillo:1992}
{Grillo}, F., {Sciortino}, S., {Micela}, G., {Vaiana}, G.~S., \& {Harnden},
  Jr., F.~R. 1992, \apjs, 81, 795

\bibitem[{{Hadrava} \& {Harmanec}(1996)}]{hadrava:1996}
{Hadrava}, P. \& {Harmanec}, P. 1996, \aap, 315, L401

\bibitem[{{Hartkopf} {et~al.}(1992){Hartkopf}, {McAlister}, \&
  {Franz}}]{hartkopf:1992}
{Hartkopf}, W.~I., {McAlister}, H.~A., \& {Franz}, O.~G. 1992, \aj, 104, 810

\bibitem[{{Henrichs} {et~al.}(1993){Henrichs}, {Bauer}, {Hill}, {Kaper},
  {Nichols-Bohlin}, \& {Veen}}]{henrichs:1993}
{Henrichs}, H.~F., {Bauer}, F., {Hill}, G.~M., {et~al.} 1993, in IAU Colloq.
  139: New Perspectives on Stellar Pulsation and Pulsating Variable Stars, ed.
  J.~M. {Nemec} \& J.~M. {Matthews}, 186

\bibitem[{{Henrichs} {et~al.}(1998){Henrichs}, {de Jong}, {Nichols}, {Kaper},
  {Bjorkman}, {Bohlender}, {Cao}, {Gordon}, {Hill}, {Jiang}, {Kolka}, {Li},
  {Liu}, {Neff}, {O'Neill}, {Scheers}, \& {Telting}}]{henrichs:1998}
{Henrichs}, H.~F., {de Jong}, J.~A., {Nichols}, J.~S., {et~al.} 1998, in ESA
  SP-413: Ultraviolet Astrophysics Beyond the IUE Final Archive, ed.
  W.~{Wamsteker}, R.~{Gonzalez Riestra}, \& B.~{Harris}, 157

\bibitem[{{Henrichs} {et~al.}(1994){Henrichs}, {Kaper}, \&
  {Nichols}}]{henrichs:1994}
{Henrichs}, H.~F., {Kaper}, L., \& {Nichols}, J.~S. 1994, \aap, 285, 565

\bibitem[{{Henrichs} {et~al.}(2012){Henrichs}, {Kolenberg}, {Plaggenborg},
  {Marsden}, {Waite}, {Landstreet}, {Wade}, {Grunhut}, \&
  {Oksala}}]{henrichs:2012}
{Henrichs}, H.~F., {Kolenberg}, K., {Plaggenborg}, B., {et~al.} 2012, \aap,
  545, A119

\bibitem[{{Henrichs} {et~al.}(2005){Henrichs}, {Schnerr}, \& {Ten
  Kulve}}]{henrichs:2005}
{Henrichs}, H.~F., {Schnerr}, R.~S., \& {Ten Kulve}, E. 2005, in ASP Conf.
  Ser., Vol. 337: The Nature and Evolution of Disks Around Hot Stars, 114

\bibitem[{{Heynderickx} {et~al.}(1994){Heynderickx}, {Waelkens}, \&
  {Smeyers}}]{heynderickx:1994}
{Heynderickx}, D., {Waelkens}, C., \& {Smeyers}, P. 1994, \aaps, 105, 447

\bibitem[{{Hubrig} {et~al.}(2011){Hubrig}, {Ilyin}, {Sch{\"o}ller}, {Briquet},
  {Morel}, \& {De Cat}}]{hubrig:2011}
{Hubrig}, S., {Ilyin}, I., {Sch{\"o}ller}, M., {et~al.} 2011, \apjl, 726, L5

\bibitem[{{Kaper} {et~al.}(1992){Kaper}, {Henrichs}, \& {Mathias}}]{kaper:1992}
{Kaper}, L., {Henrichs}, H.~F., \& {Mathias}, P. 1992, {Decline in the Halpha
  emission strength of beta Cephei}, OHP Newsletter, February

\bibitem[{{Kaper} {et~al.}(1996){Kaper}, {Henrichs}, {Nichols}, {Snoek},
  {Volten}, \& {Zwarthoed}}]{kaper:1996}
{Kaper}, L., {Henrichs}, H.~F., {Nichols}, J.~S., {et~al.} 1996, \aaps, 116,
  257

\bibitem[{{Kaper} \& {Mathias}(1995)}]{kaper:1995}
{Kaper}, L. \& {Mathias}, P. 1995, in ASP Conf. Ser. 83: IAU Colloq. 155:
  Astrophysical Applications of Stellar Pulsation, ed. R.~S. {Stobie} \& P.~A.
  {Whitelock}, 295

\bibitem[{{Landstreet}(1982)}]{landstreet:1982}
{Landstreet}, J.~D. 1982, \apj, 258, 639

\bibitem[{{Lesh}(1968)}]{lesh:1968}
{Lesh}, J.~R. 1968, \apjs, 17, 371

\bibitem[{{Mathias} {et~al.}(1991){Mathias}, {Gillet}, \&
  {Kaper}}]{mathias:1991}
{Mathias}, P., {Gillet}, D., \& {Kaper}, L. 1991, in Rapid Variability of
  OB-stars: Nature and Diagnostic Value, ed. D.~{Baade}, 193

\bibitem[{{Mathys}(1989)}]{mathys:1989}
{Mathys}, G. 1989, Fundamentals of Cosmic Physics, 13, 143

\bibitem[{{Meynet} {et~al.}(2011){Meynet}, {Eggenberger}, \&
  {Maeder}}]{meynet:2011}
{Meynet}, G., {Eggenberger}, P., \& {Maeder}, A. 2011, \aap, 525, L11

\bibitem[{{Morel} {et~al.}(2006){Morel}, {Butler}, {Aerts}, {Neiner}, \&
  {Briquet}}]{morel:2006}
{Morel}, T., {Butler}, K., {Aerts}, C., {Neiner}, C., \& {Briquet}, M. 2006,
  \aap, 457, 651

\bibitem[{{Nardetto} {et~al.}(2011){Nardetto}, {Mourard}, {Tallon-Bosc},
  {Tallon}, {Berio}, {Chapellier}, {Bonneau}, {Chesneau}, {Mathias}, {Perraut},
  {Stee}, {Blazit}, {Clausse}, {Delaa}, {Marcotto}, {Millour}, {Roussel},
  {Spang}, {McAlister}, {Ten Brummelaar}, {Sturmann}, {Sturmann}, {Turner},
  {Farrington}, \& {Goldfinger}}]{nardetto:2011}
{Nardetto}, N., {Mourard}, D., {Tallon-Bosc}, I., {et~al.} 2011, \aap, 525, A67

\bibitem[{{Neiner} {et~al.}(2003{\natexlab{a}}){Neiner}, {Geers}, {Henrichs},
  {Floquet}, {Fr{\' e}mat}, {Hubert}, {Preuss}, \& {Wiersema}}]{neiner:2003a}
{Neiner}, C., {Geers}, V.~C., {Henrichs}, H.~F., {et~al.} 2003{\natexlab{a}},
  \aap, 406, 1019

\bibitem[{{Neiner} {et~al.}(2003{\natexlab{b}}){Neiner}, {Hubert}, {Fr{\'
  e}mat}, {Floquet}, {Jankov}, {Preuss}, {Henrichs}, \& {Zorec}}]{neiner:2003b}
{Neiner}, C., {Hubert}, A.-M., {Fr{\' e}mat}, Y., {et~al.} 2003{\natexlab{b}},
  \aap, 409, 275

\bibitem[{{Nieva} \& {Przybilla}(2012)}]{nieva:2012}
{Nieva}, M.-F. \& {Przybilla}, N. 2012, \aap, 539, A143

\bibitem[{{Panek} \& {Savage}(1976)}]{panek:1976}
{Panek}, R.~J. \& {Savage}, B.~D. 1976, \apj, 206, 167

\bibitem[{{Pan'ko} \& {Tarasov}(1997)}]{panko:1997}
{Pan'ko}, E.~A. \& {Tarasov}, A.~E. 1997, Astronomy Letters, 23, 545

\bibitem[{{Petit} {et~al.}(2013){Petit}, {Owocki}, {Wade}, {Cohen},
  {Sundqvist}, {Gagn{\'e}}, {Ma{\'{\i}}z Apell{\'a}niz}, {Oksala}, {Bohlender},
  {Rivinius}, {Henrichs}, {Alecian}, {Townsend}, {ud-Doula}, \& {MiMeS
  Collaboration}}]{petit:2013}
{Petit}, V., {Owocki}, S.~P., {Wade}, G.~A., {et~al.} 2013, \mnras, 429, 398

\bibitem[{{Pigulski} \& {Boratyn}(1992)}]{pigulski:1992}
{Pigulski}, A. \& {Boratyn}, D.~A. 1992, \aap, 253, 178

\bibitem[{{Preston}(1967)}]{preston:1967}
{Preston}, G.~W. 1967, \apj, 150, 547

\bibitem[{{Rachkovskaya}(1990)}]{rachkovskaya:1990}
{Rachkovskaya}, T.~M. 1990, Bulletin Crimean Astrophysical Observatory, 82, 1

\bibitem[{{Remie} \& {Lamers}(1982)}]{remie:1982}
{Remie}, H. \& {Lamers}, H.~J.~G.~L.~M. 1982, \aap, 105, 85

\bibitem[{{Rudy} \& {Kemp}(1978)}]{rudy:1978}
{Rudy}, R.~J. \& {Kemp}, J.~C. 1978, \mnras, 183, 595

\bibitem[{{Schnerr} {et~al.}(2006{\natexlab{a}}){Schnerr}, {Henrichs},
  {Oudmaijer}, \& {Telting}}]{schnerr:2006e}
{Schnerr}, R.~S., {Henrichs}, H.~F., {Oudmaijer}, R.~D., \& {Telting}, J.~H.
  2006{\natexlab{a}}, \aap, {459}, L21

\bibitem[{{Schnerr} {et~al.}(2007){Schnerr}, {Henrichs}, {Owocki}, {Ud-Doula},
  \& {Townsend}}]{Schnerr:2007}
{Schnerr}, R.~S., {Henrichs}, H.~F., {Owocki}, S.~P., {Ud-Doula}, A., \&
  {Townsend}, R.~H.~D. 2007, in Astronomical Society of the Pacific Conference
  Series, Vol. 361, Active OB-Stars: Laboratories for Stellare and
  Circumstellar Physics, ed. A.~T. {Okazaki}, S.~P. {Owocki}, \& S.~{Stefl},
  488

\bibitem[{{Schnerr} {et~al.}(2006{\natexlab{b}}){Schnerr}, {Verdugo},
  {Henrichs}, \& {Neiner}}]{schnerr:2006a}
{Schnerr}, R.~S., {Verdugo}, E., {Henrichs}, H.~F., \& {Neiner}, C.
  2006{\natexlab{b}}, \aap, 452, 969

\bibitem[{{Schrijvers} {et~al.}(1997){Schrijvers}, {Telting}, {Aerts},
  {Ruymaekers}, \& {Henrichs}}]{schrijvers:1997}
{Schrijvers}, C., {Telting}, J.~H., {Aerts}, C., {Ruymaekers}, E., \&
  {Henrichs}, H.~F. 1997, \aaps, 121, 343

\bibitem[{{Schwarzschild}(1950)}]{schwarzschild:1950}
{Schwarzschild}, M. 1950, \apj, 112, 222

\bibitem[{{Shibahashi} \& {Aerts}(2000)}]{shibahashi:2000}
{Shibahashi}, H. \& {Aerts}, C. 2000, \apjl, 531, L143

\bibitem[{{Shore}(1987)}]{shore:1987a}
{Shore}, S.~N. 1987, \aj, 94, 731

\bibitem[{{Shore} {et~al.}(1987){Shore}, {Brown}, \& {Sonneborn}}]{shore:1987b}
{Shore}, S.~N., {Brown}, D.~N., \& {Sonneborn}, G. 1987, \aj, 94, 737

\bibitem[{{Silvester} {et~al.}(2009){Silvester}, {Neiner}, {Henrichs}, {Wade},
  {Petit}, {Alecian}, {Huat}, {Martayan}, {Power}, \& {Thizy}}]{silvester:2009}
{Silvester}, J., {Neiner}, C., {Henrichs}, H.~F., {et~al.} 2009, \mnras, 398,
  1505

\bibitem[{{Telting} {et~al.}(1997){Telting}, {Aerts}, \&
  {Mathias}}]{telting:1997}
{Telting}, J.~H., {Aerts}, C., \& {Mathias}, P. 1997, \aap, 322, 493

\bibitem[{{Townsend} {et~al.}(2005){Townsend}, {Owocki}, \&
  {Groote}}]{townsend:2005}
{Townsend}, R.~H.~D., {Owocki}, S.~P., \& {Groote}, D. 2005, \apjl, 630, L81

\bibitem[{{ud-Doula} \& {Owocki}(2002)}]{ud-doula:2002}
{ud-Doula}, A. \& {Owocki}, S.~P. 2002, \apj, 576, 413

\bibitem[{{Ud-Doula} {et~al.}(2008){Ud-Doula}, {Owocki}, \&
  {Townsend}}]{ud-Doula:2008}
{Ud-Doula}, A., {Owocki}, S.~P., \& {Townsend}, R.~H.~D. 2008, \mnras, 385, 97

\bibitem[{{Ud-Doula} {et~al.}(2009){Ud-Doula}, {Owocki}, \&
  {Townsend}}]{ud-doula:2009}
{Ud-Doula}, A., {Owocki}, S.~P., \& {Townsend}, R.~H.~D. 2009, \mnras, 392,
  1022

\bibitem[{{van Leeuwen}(2007)}]{vanleeuwen:2007}
{van Leeuwen}, F. 2007, \aap, 474

\bibitem[{{Wade} {et~al.}(2006){Wade}, {Auri{\`e}re}, {Bagnulo}, {Donati},
  {Johnson}, {Landstreet}, {Ligni{\`e}res}, {Marsden}, {Monin}, {Mouillet},
  {Paletou}, {Petit}, {Toqu{\'e}}, {Alecian}, \& {Folsom}}]{wade:2006}
{Wade}, G.~A., {Auri{\`e}re}, M., {Bagnulo}, S., {et~al.} 2006, \aap, 451, 293

\bibitem[{{Wade} {et~al.}(1997){Wade}, {Bohlender}, {Brown}, {Elkin},
  {Landstreet}, \& {Romanyuk}}]{wade:1997}
{Wade}, G.~A., {Bohlender}, D.~A., {Brown}, D.~N., {et~al.} 1997, \aap, 320,
  172

\bibitem[{{Wade} {et~al.}(2000){Wade}, {Donati}, {Landstreet}, \&
  {Shorlin}}]{wade:2000}
{Wade}, G.~A., {Donati}, J.-F., {Landstreet}, J.~D., \& {Shorlin}, S.~L.~S.
  2000, \mnras, 313, 851

\bibitem[{{Wade} {et~al.}(2007){Wade}, {Silvester}, {Bale}, {Johnson}, {Power},
  {Auri{\`e}re}, {Ligni{\'e}res}, {Dintrans}, {Donati}, {Bon Hoa}, {Mouillet},
  {Naseri}, {Paletou}, {Petit}, {Rincon}, {Toque}, {Bagnulo}, {Folsom},
  {Landstreet}, {Gruberbauer}, {Lueftinger}, {Jeffers}, {L{\`e}bre}, \&
  {Marsden}}]{wade:2007}
{Wade}, G.~A., {Silvester}, J., {Bale}, K., {et~al.} 2007, ArXiv e-prints

\bibitem[{{Wheelwright} {et~al.}(2009){Wheelwright}, {Oudmaijer}, \&
  {Schnerr}}]{wheelwright:2009}
{Wheelwright}, H.~E., {Oudmaijer}, R.~D., \& {Schnerr}, R.~S. 2009, \aap, 497,
  487

\end{thebibliography}

\end{document}